%Paper: nucl-th/9409016
%From: Masayuki Matsuzaki <matsuza@fukuoka-edu.ac.jp>
%Date: Mon, 19 Sep 1994 14:58:23 +0900

%%%%%%%%%%%%%%%%%%%%%%%%%%%%%%%%%%%%%%%%%%%%%%%%%%%%%%%%%%%%%%%%%%%%%
%
%\immediate\write16{%
%         This is ApTEX  Ver 1.1 (c) Sept/18/1989 M.Nagasawa}
%
%     This is PHYZZX macro package extended by M.Nagasawa 1987
%     This version of PHYZZX should be used with Version 1.0 of TEX
%
\catcode`@=11 % This allows us to modify PLAIN macros.
%
%      Combined "efont.lib,format.lib,page.lib,figspc.lib"
%       with small modification for only figure-spacing by YRS
%                                         Feb. 1990
%
%%%%%%%%%%%%%%%%%%%%%%%%%%%%%%%%%%%%%%%%%%%%%%%%%%%%%%%%%%%%%%%%%%%%%
%
%    fonts
%
%%%%%%%%%%%%%%%%%%%%%%%%%%%%%%%%%%%%%%%%%%%%%%%%%%%%%%%%%%%%%%%%%%%%%
%

\font\fourteenrm=cmr10 scaled\magstep2
\font\twelverm=cmr10 scaled\magstep1
\font\elevenrm=cmr10 scaled\magstephalf
\font\ninerm=cmr9

\font\sevenrm=cmr7
\font\sixrm=cmr6
\font\fiverm=cmr5

\font\fourteenbf=cmbx10 scaled\magstep2
\font\twelvebf=cmbx10 scaled\magstep1
\font\elevenbf=cmbx10 scaled\magstephalf
\font\ninebf=cmbx9
\font\sixbf=cmbx6
\font\twentyi=cmmi12 scaled\magstep3	   \skewchar\twentyi='177
\font\seventeeni=cmmi12 scaled\magstep2	   \skewchar\seventeeni='177
\font\fourteeni=cmmi10 scaled\magstep2	   \skewchar\fourteeni='177
\font\twelvei=cmmi10 scaled\magstep1	   \skewchar\twelvei='177
\font\eleveni=cmmi10 scaled\magstephalf    \skewchar\eleveni='177
\font\ninei=cmmi9			   \skewchar\ninei='177
\font\sixi=cmmi6			   \skewchar\sixi='177
\font\twentysy=cmsy10 scaled\magstep4      \skewchar\twentysy='60
\font\seventeensy=cmsy10 scaled\magstep3   \skewchar\seventeensy='60
\font\fourteensy=cmsy10 scaled\magstep2	   \skewchar\fourteensy='60
\font\twelvesy=cmsy10 scaled\magstep1	   \skewchar\twelvesy='60
\font\elevensy=cmsy10 scaled\magstephalf   \skewchar\elevensy='60
\font\ninesy=cmsy9			   \skewchar\ninesy='60
\font\sixsy=cmsy6			   \skewchar\sixsy='60

\font\fourteenex=cmex10 scaled\magstep2
\font\twelveex=cmex10 scaled\magstep1
\font\elevenex=cmex10 scaled\magstephalf

\font\fourteensl=cmsl10 scaled\magstep2
\font\twelvesl=cmsl10 scaled\magstep1
\font\elevensl=cmsl10 scaled\magstephalf
\font\ninesl=cmsl9

\font\fourteenit=cmti10 scaled\magstep2
\font\twelveit=cmti10 scaled\magstep1
\font\elevenit=cmti10 scaled\magstephalf
\font\twelvett=cmtt10 scaled\magstep1
\font\eleventt=cmtt10 scaled\magstephalf

\font\twelvecp=cmcsc10 scaled\magstep1
\font\elevencp=cmcsc10 scaled\magstephalf
\font\tencp=cmcsc10
\newfam\cpfam
\newcount\f@ntkey \f@ntkey=0
\def\samef@nt{\relax \ifcase\f@ntkey \rm \or\oldstyle \or\or
	 \or\it \or\sl \or\bf \or\tt \or\caps \fi}
\def\fourteenpoint{\relax%
        \textfont0=\fourteenrm \scriptfont0=\tenrm%
        \scriptscriptfont0=\sevenrm%
    \def\rm{\fam0 \fourteenrm \f@ntkey=0 }\relax%
        \textfont1=\fourteeni \scriptfont1=\teni
        \scriptscriptfont1=\seveni%
    \def\oldstyle{\fam1 \fourteeni\f@ntkey=1 }\relax%
        \textfont2=\fourteensy \scriptfont2=\tensy%
        \scriptscriptfont2=\sevensy%
        \textfont3=\fourteenex \scriptfont3=\fourteenex%
        \scriptscriptfont3=\fourteenex%
    \def\it{\fam\itfam \fourteenit\f@ntkey=4 }%
        \textfont\itfam=\fourteenit%
    \def\sl{\fam\slfam \fourteensl\f@ntkey=5 }%
        \textfont\slfam=\fourteensl \scriptfont\slfam=\tensl%
    \def\bf{\fam\bffam \fourteenbf\f@ntkey=6 }
        \textfont\bffam=\fourteenbf
        \scriptfont\bffam=\tenbf \scriptscriptfont\bffam=\sevenbf
    \def\tt{\fam\ttfam \twelvett \f@ntkey=7 }%
        \textfont\ttfam=\twelvett%
        \h@big=11.9\p@{} \h@Big=16.1\p@{}%
        \h@bigg=20.3\p@{} \h@Bigg=24.5\p@{}%
    \def\caps{\fam\cpfam \twelvecp \f@ntkey=8 }%
        \textfont\cpfam=\twelvecp%
    \setbox\strutbox=\hbox{\vrule height12pt depth5pt width\z@ }%
    \samef@nt }
\def\twelvepoint{\relax
        \textfont0=\twelverm \scriptfont0=\ninerm
        \scriptscriptfont0=\sixrm
    \def\rm{\fam0 \twelverm \f@ntkey=0 }\relax
        \textfont1=\twelvei \scriptfont1=\ninei
        \scriptscriptfont1=\sixi
    \def\oldstyle{\fam1 \twelvei\f@ntkey=1 }\relax
        \textfont2=\twelvesy \scriptfont2=\ninesy
        \scriptscriptfont2=\sixsy
        \textfont3=\twelveex \scriptfont3=\twelveex
        \scriptscriptfont3=\twelveex
    \def\it{\fam\itfam \twelveit \f@ntkey=4 }
        \textfont\itfam=\twelveit
    \def\sl{\fam\slfam \twelvesl \f@ntkey=5 }
        \textfont\slfam=\twelvesl \scriptfont\slfam=\ninesl
    \def\bf{\fam\bffam \twelvebf \f@ntkey=6 \relax}
        \textfont\bffam=\twelvebf \scriptfont\bffam=\ninebf
        \scriptscriptfont\bffam=\sixbf
    \def\tt{\fam\ttfam \twelvett \f@ntkey=7 }
        \textfont\ttfam=\twelvett
        \h@big=10.2\p@{} \h@Big=13.8\p@{}
        \h@bigg=17.4\p@{} \h@Bigg=21.0\p@{}
    \def\caps{\fam\cpfam \twelvecp \f@ntkey=8 }
        \textfont\cpfam=\twelvecp
    \setbox\strutbox=\hbox{\vrule height 10pt depth 4pt width\z@ }
    \samef@nt }
\def\elevenpoint{\relax
        \textfont0=\elevenrm \scriptfont0=\sevenrm
        \scriptscriptfont0=\fiverm
    \def\rm{\fam0 \elevenrm \f@ntkey=0 }\relax
        \textfont1=\eleveni \scriptfont1=\seveni
        \scriptscriptfont1=\fivei
    \def\oldstyle{\fam1 \eleveni \f@ntkey=1 }\relax
        \textfont2=\elevensy \scriptfont2=\sevensy
        \scriptscriptfont2=\fivesy
        \textfont3=\elevenex \scriptfont3=\elevenex
        \scriptscriptfont3=\elevenex
   \def\it{\fam\itfam \elevenit \f@ntkey=4 }\textfont\itfam=\elevenit
   \def\sl{\fam\slfam \elevensl \f@ntkey=5 }\textfont\slfam=\elevensl
   \def\bf{\fam\bffam \elevenbf \f@ntkey=6 }
        \textfont\bffam=\elevenbf \scriptfont\bffam=\sevenbf
        \scriptscriptfont\bffam=\fivebf
 \def\tt{\fam\ttfam \eleventt \f@ntkey=7 }\textfont\ttfam=\eleventt
 \def\caps{\fam\cpfam \elevencp \f@ntkey=8 }\textfont\cpfam=\elevencp
    \setbox\strutbox=\hbox{\vrule height 9.3pt depth 3.8pt width\z@ }
    \samef@nt }
\def\tenpoint{\relax
        \textfont0=\tenrm \scriptfont0=\sevenrm
        \scriptscriptfont0=\fiverm
    \def\rm{\fam0 \tenrm \f@ntkey=0 }\relax
        \textfont1=\teni \scriptfont1=\seveni
        \scriptscriptfont1=\fivei
    \def\oldstyle{\fam1 \teni \f@ntkey=1 }\relax
        \textfont2=\tensy \scriptfont2=\sevensy
        \scriptscriptfont2=\fivesy
        \textfont3=\tenex \scriptfont3=\tenex
        \scriptscriptfont3=\tenex
    \def\it{\fam\itfam \tenit \f@ntkey=4 }\textfont\itfam=\tenit
    \def\sl{\fam\slfam \tensl \f@ntkey=5 }\textfont\slfam=\tensl
    \def\bf{\fam\bffam \tenbf \f@ntkey=6 }
        \textfont\bffam=\tenbf \scriptfont\bffam=\sevenbf
        \scriptscriptfont\bffam=\fivebf
    \def\tt{\fam\ttfam \tentt \f@ntkey=7 }\textfont\ttfam=\tentt
    \def\caps{\fam\cpfam \tencp \f@ntkey=8 }\textfont\cpfam=\tencp
    \setbox\strutbox=\hbox{\vrule height 8.5pt depth 3.5pt width\z@ }
    \samef@nt }
\newdimen\h@big  \h@big=8.5\p@
\newdimen\h@Big  \h@Big=11.5\p@
\newdimen\h@bigg  \h@bigg=14.5\p@
\newdimen\h@Bigg  \h@Bigg=17.5\p@
\def\big#1{{\hbox{$\left#1\vbox to\h@big{}\right.\n@space$}}}
\def\Big#1{{\hbox{$\left#1\vbox to\h@Big{}\right.\n@space$}}}
\def\bigg#1{{\hbox{$\left#1\vbox to\h@bigg{}\right.\n@space$}}}
\def\Bigg#1{{\hbox{$\left#1\vbox to\h@Bigg{}\right.\n@space$}}}
\def\blank{ }
%
%%%%%%%%%%%%%%%%%%%%%%%%%%%%%%%%%%%%%%%%%%%%%%%%%%%%%%%%%%%%%%%%%%%%%
%
%     basic spacing parameters.
%
%%%%%%%%%%%%%%%%%%%%%%%%%%%%%%%%%%%%%%%%%%%%%%%%%%%%%%%%%%%%%%%%%%%%%
%
\tolerance=1600
\hfuzz=1pt \vfuzz=12pt
\parindent=22pt
\def\myhyphen{\tolerance=2000 \pretolerance=3600
       \hbadness=3600 \hyphenpenalty=500 \exhyphenpenalty=500
       \interlinepenalty=1000 \predisplaypenalty=9000
       \postdisplaypenalty=500 \binoppenalty=700 \relpenalty=500}
\newif\ifsingl@
\newif\ifdoubl@
\def\singlespace{\singl@true\doubl@false\spaces@t}
\def\doublespace{\singl@false\doubl@true\spaces@t}
\def\normalspace{\singl@false\doubl@false\spaces@t}
\newcount\fontsize \fontsize=12
\newif\iftwelv@ \twelv@true
\def\Tenpoint{\tenpoint \fontsize=10 \spaces@t}
\def\Elevenpoint{\elevenpoint \fontsize=11 \spaces@t}
\def\Twelvepoint{\twelvepoint \fontsize=12 \spaces@t}
\def\spaces@t{\relax%
    \ifnum\fontsize=12%
       \ifsingl@\subspaces@t3:4;\else\subspaces@t7:8;\fi\fi%
    \ifnum\fontsize=11%
       \ifsingl@\subspaces@t5:7;\else\subspaces@t6:7;\fi\fi%
    \ifnum\fontsize=10%
       \ifsingl@\subspaces@t3:5;\else\subspaces@t4:5;\fi\fi%
    \ifdoubl@ \multiply\baselineskip by 5%
              \divide\baselineskip by 4 \fi}
\def\subspaces@t#1:#2;{%
    \baselineskip=\normalbaselineskip%
    \multiply\baselineskip by #1 \divide\baselineskip by #2%
    \lineskip=\normallineskip%
    \multiply\lineskip by #1 \divide\lineskip by #2%
    \lineskiplimit=\normallineskiplimit%
    \multiply\lineskiplimit by #1 \divide\lineskiplimit by #2%
    \parskip=\normalparskip%
    \multiply\parskip by #1 \divide\parskip by #2%
    \abovedisplayskip=\normdisplayskip%
    \multiply\abovedisplayskip by #1 \divide\abovedisplayskip by #2%
    \belowdisplayskip=\abovedisplayskip%
    \abovedisplayshortskip=\normaldispshortskip%
    \multiply\abovedisplayshortskip by #1%
    \divide\abovedisplayshortskip by #2%
    \belowdisplayshortskip=\abovedisplayshortskip%
    \advance\belowdisplayshortskip by \belowdisplayskip%
    \divide\belowdisplayshortskip by 2%
    \smallskipamount=\skipamount%
    \multiply\smallskipamount by #1 \divide\smallskipamount by #2%
    \medskipamount=\smallskipamount \multiply\medskipamount by 2%
    \bigskipamount=\smallskipamount \multiply\bigskipamount by 4}

\def\figbaselines{\baselineskip=10.0pt  \lineskip=1.0pt
                  \lineskiplimit=1.0pt}
\def\normalbaselines{\baselineskip=\normalbaselineskip%
    \lineskip=\normallineskip \lineskiplimit=\normallineskip%
    \ifnum\fontsize=10
       \multiply\baselineskip by 4 \divide\baselineskip by 5%
       \multiply\lineskiplimit by 4 \divide\lineskiplimit by 5%
       \multiply\lineskip by 4 \divide\lineskip by 5 \fi
   \ifnum\fontsize=11
       \multiply\baselineskip by 6 \divide\baselineskip by 7%
       \multiply\lineskiplimit by 6 \divide\lineskiplimit by 7%
       \multiply\lineskip by 6 \divide\lineskip by 7 \fi }
\def\spacecheck#1{\dimen@=\pagegoal\advance\dimen@ by -\pagetotal
    \ifdim\dimen@<#1 \ifdim\dimen@>0pt \vfil\break \fi\fi}
\normalbaselineskip=20pt plus 0.2pt minus 0.1pt
\normallineskip=1.5pt plus 0.1pt minus 0.1pt
\normallineskiplimit=1.5pt
\newskip\normdisplayskip    \normdisplayskip=15pt plus 4pt minus 8pt
\newskip\normaldispshortskip \normaldispshortskip=6pt plus 5pt
\newskip\normalparskip      \normalparskip=2pt plus 0.2pt minus 0.2pt
\newskip\skipamount          \skipamount=5pt plus 2pt minus 1.5pt
\newskip\headskipamount	     \headskipamount=0pt plus 3pt minus 3pt
\newskip\frontpageskip       \frontpageskip=0pt plus .1fil
\def\oneskip{\vskip\baselineskip}
\def\headskip{\vskip\headskipamount}
\newdimen\normalhsize
\newdimen\halfhsize
\newdimen\normalvsize
\newdimen\halfvsize     \halfvsize=115mm
%
%%%%%%%%%%%%%%%%%%%%%%%%%%%%%%%%%%%%%%%%%%%%%%%%%%%%%%%%%%%%%%%%%%%%%
%
%   page output routines.
%
%%%%%%%%%%%%%%%%%%%%%%%%%%%%%%%%%%%%%%%%%%%%%%%%%%%%%%%%%%%%%%%%%%%%%
%
\def\backup{\penalty 100%
      \splittopskip=0pt \splitmaxdepth=\maxdimen \floatingpenalty=0%
       \begingroup \global\setbox0=\vbox{\unvbox255}%
            \dimen@=\ht0 \advance\dimen@ by\dp0%
            \advance\dimen@ by12pt \advance\dimen@ by\pagetotal%
            \ifdim\dimen@>\pagegoal%
            \p@gefalse \fi%
            \endgroup}
\newif\ifretry  \retryfalse
\def\plainoutput{\ifretry\global\retryfalse%
               \global\advance\vsize by\ht0%
         \shipout\vbox{\makeheadline\unvbox255\unvbox0\makefootline}%
         \else\shipout\vbox{\pageall}\fi%
              \advancepageno%
              \ifnum\outputpenalty>-20000\else\dosupereject\fi}
\def\pageall{\makeheadline%
             \vbox to\vsize{\boxmaxdepth=\maxdepth \pagecontents}%
             \makefootline}
\newcount\hangafterA
\newskip\hangindentA
\newcount\prevgrafA
\def\par{\hangafterA=\the\hangafter \hangindentA=\the\hangindent%
   \endgraf%
   \ifnum\prevgraf<-\hangafterA \prevgrafA=\the\prevgraf%
   \hangafter=\the\hangafterA \hangindent=\the\hangindentA%
   \indent\prevgraf=\the\prevgrafA\fi}%

\def\nextline{\ifnum\referencecount<-1 \endgraf \hskip7.0ex
              \else \unskip\nobreak\hskip\parfillskip\break \fi}
\countdef\pagenumber=1  \pagenumber=1
\newcount\chapternumber \chapternumber=0
\newcount\sectionnumber	\sectionnumber=0
\newcount\subsectionnumber \subsectionnumber=0
\newcount\equanumber \equanumber=0

\newif\iffrontpage
\def\FRONTPAGE{\ifvoid255\else\bigskip\vskip\frontpageskip
               \penalty-2000\fi
               \global\frontpagetrue \global\lastf@@t=0
               \global\footsymbolcount=0}
\newif\ifp@genum \p@genumtrue
\def\nopagenumbers{\p@genumfalse}
\def\pagenumbers{\p@genumtrue}
\def\pagecontents{%
    \ifvoid\topins\else\unvbox\topins\vskip\skip\topins\fi
    \dimen@=\dp255 \unvbox255
    \ifvoid\footins\else
       \ifnum\footsymbolcount<0\else\vskip\skip\footins \footrule\fi
       \unvbox\footins
    \fi%
    \ifr@ggedbottom \kern-\dimen@ \vfil \fi}
\def\advancepageno{\global\advance\pageno by 1
    \ifnum\pagenumber<0 \global\advance\pagenumber by -1
    \else\global\advance\pagenumber by 1 \fi
    \global\frontpagefalse}
\def\folio{\ifnum\pagenumber<0 \romannumeral-\pagenumber
	   \else \number\pagenumber \fi}
\def\fullline{\hbox to\normalhsize}
\newtoks\oddheadline  \oddheadline={ }
\newtoks\evenheadline \evenheadline={ }
\newtoks\paperheadline
\paperheadline={\hss\iffrontpage\else
             \ifodd\pageno\the\oddheadline \else\the\evenheadline \fi
             \hss \fi}
\headline={\the\paperheadline}
\def\makeheadline{\vbox to0pt{\skip@=\topskip
    \advance\skip@ by -12pt \advance\skip@ by -2\normalbaselineskip
    \vskip\skip@ \fullline{\vbox to12pt{}\the\headline}\vss}
    \nointerlineskip}
%
% *** [page-footer] killed from original ***
%
%\newtoks\paperfootline
%\paperfootline={\hss\iffrontpage
%                \else\ifp@genum\twelverm-- \folio \ --\hss\fi\fi}
%\footline={\the\paperfootline}
%\def\makefootline{\baselineskip=1.5\normalbaselineskip
%    \fullline{\the\footline}}
%
\def\attach#1{\space@ver{\strut^{\mkern 2mu #1}}\@sf}
%
%%%%%%%%%%%%%%%%%%%%%%%%%%%%%%%%%%%%%%%%%%%%%%%%%%%%%%%%%%%%%%%%%%%%%
%
%   footnotes
%
%%%%%%%%%%%%%%%%%%%%%%%%%%%%%%%%%%%%%%%%%%%%%%%%%%%%%%%%%%%%%%%%%%%%%
%
\def\footnote#1{\attach{#1}\vfootnote{#1}}

\def\footrule{\dimen@=\prevdepth \nointerlineskip
    \vbox to0pt{\vskip-0.25\baselineskip \hrule width0.35\hsize \vss}
    \prevdepth=\dimen@ }
\newtoks\foottokens     \foottokens={\tenpoint\singlespace}
\newdimen\footindent    \footindent=24pt
\def\vfootnote#1{\insert\footins\bgroup  \the\foottokens
    \interlinepenalty=5000 \floatingpenalty=20000
    \splittopskip=\ht\strutbox \splitmaxdepth=\dp\strutbox
    \leftskip=\footindent \rightskip=\z@skip
    \parindent=0.5\footindent \parfillskip=0pt plus 1fil
    \spaceskip=\z@skip \xspaceskip=\z@skip
    \textindent{$ #1 $}\footstrut\futurelet\next\fo@t}
\newcount\lastf@@t	     \lastf@@t=-1
\newcount\footsymbolcount    \footsymbolcount=0
\def\fd@f#1{\xdef\footsymbol{#1}}
\let\footsymbol=\star
\def\footsymbolgen{\relax \NPsymbolgen
    \global\lastf@@t=\pageno}
\def\NPsymbolgen{%
    \ifnum\footsymbolcount<0 \global\footsymbolcount=0\fi
    {\iffrontpage \else \advance\lastf@@t by 1 \fi
    \ifnum\lastf@@t<\pageno \global\footsymbolcount=0
    \else \global\advance\footsymbolcount by 1 \fi }
    \ifcase\footsymbolcount
    \fd@f\star\or \fd@f\dagger\or \fd@f\ast\or
    \fd@f\ddagger\or \fd@f\natural\or \fd@f\diamond\or
    \fd@f\bullet\or \fd@f\nabla\else
    \fd@f\dagger\global\footsymbolcount=0 \fi}
%
%%%%%%%%%%%%%%%%%%%%%%%%%%%%%%%%%%%%%%%%%%%%%%%%%%%%%%%%%%%%%%%%%%%%%
%
%    figure captions
%
%%%%%%%%%%%%%%%%%%%%%%%%%%%%%%%%%%%%%%%%%%%%%%%%%%%%%%%%%%%%%%%%%%%%%
%
\newcount\figurecount	  \figurecount=0
\def\fignum#1{\global\advance\figurecount by 1%
    \xdef#1{\the\figurecount}\message{[Fig.\the\figurecount]}}

\def\Figleft/#1/#2{\fignum\?\expandafter\figleft/\?/#1/{#2}}
\def\Figright/#1/#2{\fignum\?\expandafter\figright/\?/#1/{#2}}
\def\Figcenter/#1/#2{\fignum\?\expandafter\figcenter/\?/#1/{#2}}
\def\Figtop/#1/#2{\fignum\?\expandafter\figtop/\?/#1/{#2}}
\def\Figfoot/#1/#2{\fignum\?\expandafter\figfoot/\?/#1/{#2}}
\def\Figpage/#1{\fignum\?\expandafter\figpage/\?/{#1}}
\def\Figscenter/#1/#2#3{\fignum\?\fignum\!%
                       \figscenter/\?\!/#1/{#2}{#3}}
\def\Figspage/#1#2{\fignum\?\fignum\!\figspage/\?\!/{#1}{#2}}
\def\Figstop/#1/#2#3{\fignum\?\fignum\!\figstop/\?\!/#1/{#2}{#3}}
\def\Figsfoot/#1/#2#3{\fignum\?\fignum\!\figsfoot/\?\!/#1/{#2}{#3}}
\def\FIGleft#1/#2/#3{\fignum#1\figleft/#1/#2/{#3}}
\def\FIGright#1/#2/#3{\fignum#1\figright/#1/#2/{#3}}
\def\FIGcenter#1/#2/#3{\fignum#1\figcenter/#1/#2/{#3}}
\def\FIGtop#1/#2/#3{\fignum#1\figtop/#1/#2/{#3}}
\def\FIGfoot#1/#2/#3{\fignum#1\figfoot/#1/#2/{#3}}
\def\FIGpage#1/#2{\fignum#1\figpage/#1/{#2}}
\def\FIGScenter#1#2/#3/#4#5{\fignum#1\fignum#2%
                       \figscenter/#1#2/#3/{#4}{#5}}
\def\FIGSpage#1#2/#3#4{\fignum#1\fignum#2\figspage/#1#2/{#3}{#4}}
\def\FIGStop#1#2/#3/#4#5{\fignum#1\fignum#2\figstop/#1#2/#3/{#4}{#5}}
\def\FIGSfoot#1#2/#3/#4#5{\fignum#1\fignum#2%
                       \figsfoot/#1#2/#3/{#4}{#5}}
%
%%%%%%%%%%%%%%%%%%%%%%%%%%%%%%%%%%%%%%%%%%%%%%%%%%%%%%%%%%%%%%%%%%%%%
%
\catcode`@=11 % This allows us to modify PLAIN macros.
%
%    figure space
%
%%%%%%%%%%%%%%%%%%%%%%%%%%%%%%%%%%%%%%%%%%%%%%%%%%%%%%%%%%%%%%%%%%%%%
%
%--------------------------------------------------------------------
\newdimen\shiftwindow \shiftwindow=0pt
\newcount\hhfigcap
\newcount\mmfigcap
\newdimen\figcapht
\newdimen\figcapsft
\newif\if@mid
\newif\ifp@ge
\def\figcaptxt/#1/#2{\tenpoint\singlespace\figbaselines
                      Fig.#1.\ #2 \endgraf\oneskip}
\def\figcapbox/#1/#2/#3{\vbox{\endgraf%
              \parskip=0mm \parindent=0pt
              \if#1C \hangia=0.054\hsize \hangic=0.092\hsize%
                     \hangib=0.892\hsize \hangid=0.854\hsize%
              \else  \hangib=0.414\hsize \hangid=\hangib%
                     \advance\hangib by 6mm%
                     \hangia=2mm \hangic=8mm
                \if#1R \advance\hangia by 0.53\hsize
                       \advance\hangic by 0.53\hsize \fi \fi
              \parshape=2 \hangia \hangib \hangic \hangid
             {\figcaptxt/#2/{#3}}
              }}
\def\graphfile{\blank}
\def\graphcontent{\if\graphfile\blank\vss%
                  \else\vss\input\graphfile\vss\fi}
\def\figcolumn/#1/#2/#3/#4{\myhyphen%
              \FIGPAR#1{#3}%
              \global\figcapht=\baselineskip%
              \ifnum\mmfigcap>0%
                \global\multiply\figcapht by\mmfigcap%
              \else%
                \global\figcapsft=\baselineskip%
                \global\multiply\figcapsft by\mmfigcap%
                \vskip\figcapsft%
                \global\multiply\figcapht by#3%
              \fi%
              \vbox to\figcapht{\if#1R\shiftwindow=0.42\hsize\fi%
                         \graphcontent\figcapbox/#1/#2/{#4}}%
              \vskip-\figcapht%
              \ifnum\mmfigcap>0\else\vskip-\figcapsft\fi%
              \let\graphfile=\blank \shiftwindow=0pt}%
%--------------------------------------------------------------------
\newcount\figheadstart  \figheadstart=3
\newcount\hangbefore
\def\FIGPAR#1#2{%
    \hangid=\hsize \divide\hangid by 2 \if#1R \hangid=-\hangid \fi%
    \edef\next{\hangafter=\figheadstart \hangindent=\hangid}%
    \ifnum\figheadstart>0 \next \fi%
    \endgraf\hangbefore=\the\prevgraf \parskip=0pt \next%
    \ifnum\hangbefore>\hangafter%
      \advance\hangbefore by-\figheadstart% \nobreak%
    \else \hangbefore=0 \penalty-3000 \fi%
    \hangafter=-#2 \advance\hangafter by\hangbefore%
    \message{(\the\figheadstart,\the\hangafter,\the\hangbefore)}%
    \ifnum\hangafter>-1 \hangafter=999 \global\mmfigcap=-#2%
    \else \global\mmfigcap=-\the\hangafter \fi}
%--------------------------------------------------------------------
\newskip\hangia
\newskip\hangib
\newskip\hangic
\newskip\hangid
\newskip\hfigcap
\newskip\mfigcap

\def\figtop/#1/#2/#3{%
            \@midfalse\p@gefalse\begingroup%
            \mmfigcap=#2 \global\figcapht=\normalbaselineskip%
            \global\multiply\figcapht by\mmfigcap%
            \setbox0=\vbox%
              to\figcapht{\graphcontent\figcapbox/C/#1/{#3}}%
            \insert\topins{\penalty100 \floatingpenalty=0%
            \splittopskip=0pt \splitmaxdepth=\maxdimen%
            \box0\break}\endgroup\let\graphfile=\blank}
\def\figfoot/#1/#2/#3{%
            \@midfalse\p@gefalse\begingroup%
            \mmfigcap=#2 \global\figcapht=\normalbaselineskip%
            \global\multiply\figcapht by\mmfigcap%
            \setbox0=\vbox%
              to\figcapht{\graphcontent\figcapbox/C/#1/{#3}}%
            \global\footsymbolcount=-1%
            \insert\footins{\penalty100 \floatingpenalty=0%
            \splittopskip=0pt \splitmaxdepth=\maxdimen%
            \box0 \break}%
            \endgroup\let\graphfile=\blank}%
\def\figstop/#1#2/#3/#4#5{%
            \@midfalse\p@gefalse\begingroup%
            \mmfigcap=#3 \global\figcapht=\baselineskip%
            \global\multiply\figcapht by\mmfigcap%
               \setbox0=\hbox to\hsize{%
                \vbox to\figcapht{\graphcontent\figcapbox/L/#1/{#4}}%
                \hfill \vbox to\figcapht{\vss\figcapbox/L/#2/{#5}}}%
            \insert\topins{\penalty100 \floatingpenalty=0%
            \splittopskip=0pt \splitmaxdepth=\maxdimen%
            \box0 \break} \endgroup\let\graphfile=\blank}%
\def\figsfoot/#1#2/#3/#4#5{%
            \@midfalse\p@gefalse\begingroup%
            \mmfigcap=#3 \global\figcapht=\baselineskip%
            \global\multiply\figcapht by\mmfigcap%
               \setbox0=\hbox to\hsize{%
                \vbox to\figcapht{\graphcontent\figcapbox/L/#1/{#4}}%
                \hfill \vbox to\figcapht{\vss\figcapbox/L/#2/{#5}}}%
            \global\footsymbolcount=-1%
            \insert\footins{\penalty100 \floatingpenalty=0%
            \splittopskip=0pt \splitmaxdepth=\maxdimen%
            \box0 \break} \endgroup\let\graphfile=\blank}%
\def\figpage/#1/#2{%
             \@midfalse\p@getrue\begingroup%
             \setbox0=\vbox{\graphcontent\figcapbox/C/#1/{#2}}%
             \insert\topins{\penalty100 \floatingpenalty=0%
             \splittopskip=0pt \splitmaxdepth=\maxdimen \dimen@=\dp0%
             \vbox to\vsize{\vfill \unvbox0 \kern-\dimen@}}\endgroup%
             \let\graphfile=\blank%
             }
\def\figspage/#1#2/#3#4{%
             \@midfalse\p@getrue\begingroup%
             \setbox0=\hbox{%
               \vbox{\graphcontent\figcapbox/L/#1/{#3}}%
               \hskip 5mm%
               \vbox{\vss\figcapbox/L/#2/{#4}}}%
             \insert\topins{\penalty100 \floatingpenalty=0%
             \splittopskip=0pt \splitmaxdepth=\maxdimen \dimen@=\dp0%
             \vbox to\vsize{\vfill \box0 \kern-\dimen@}}\endgroup%
             \let\graphfile=\blank%
             }
\def\figleft/#1/#2/#3{\figcolumn/L/#1/#2/{#3}}
\def\figright/#1/#2/#3{\figcolumn/R/#1/#2/{#3}}
\def\figcenter/#1/#2/#3{%
              \@midtrue\begingroup
              \mmfigcap=#2 \global\figcapht=\baselineskip%
              \global\multiply\figcapht by\mmfigcap%
%%            \global\vbadness=10000
              \setbox0=\vbox
                to\figcapht{\graphcontent\figcapbox/C/#1/{#3}}
              \dimen@=\ht0
%%            \advance\dimen@ by12pt
              \advance\dimen@ by\pagetotal
              \ifdim\dimen@>\pagegoal
                   \endgraf \box0 \endgraf
              \else %% \break
                   \endgraf \box0 \endgraf\fi
              \endgroup\let\graphfile=\blank
              }
\def\figscenter/#1#2/#3/#4#5{%
               \@midtrue\begingroup
               \mmfigcap=#3 \global\figcapht=\normalbaselineskip%
               \global\multiply\figcapht by\mmfigcap%
               \global\vbadness=10000
               \setbox0=\hbox{%
                 \vbox
                   to\figcapht{\graphcontent\figcapbox/L/#1/{#4}}
                 \vbox
                  to\figcapht{\vss\figcapbox/L/#2/{#5}}}
               \if@mid \dimen@=\ht0 \advance\dimen@ by\dp0
               \advance\dimen@ by12pt \advance\dimen@ by\pagetotal
               \ifdim\dimen@>\pagegoal\@midfalse\p@gefalse\fi\fi
               \endgraf \box0 \smallbreak \endgraf \endgroup
               \let\graphfile=\blank
               }
\catcode`@=12 % at signs are no longer characters
%
%nosupertex
%noinput
%%%%%%%%%%%%%%%%%%%%%%%%%%%%%%%%%%%%%%%%%%%%%%%%%%%%%%%%%%%%%%%%%%%%%
%
% This is the 1st version of reference macros
%                                              9 Aug 1987   K Nemoto
%         no use of file for keepref          11 Aug 1987
%         checker of redefinition of keepref  23 Aug 1987
%         The latest time stamp of the source "refmacro.tex"
%                                             87-08-24 16:41
%
%    Minimum Setup : can be used independently of "PHYZZX" (YRS)
%
%%%%%%%%%%%%%%%%%%%%%%%%%%%%%%%%%%%%%%%%%%%%%%%%%%%%%%%%%%%%%%%%%%%%%
\catcode `\@=11
\let\rel@x=\relax
\let\n@expand=\relax
\def\pr@tect{\let\n@expand=\noexpand}
\let\protect=\pr@tect
\let\gl@bal=\global
%%%%%%%%%%%%%%%%%%%%%%%%%%%%%%%%%%%%%%%%%%%%%%%%%%%%%%%%%%%%%%%%%
%
% First we define the macros required for sorting reference-numbers
%
%\input sort.tex
%
% Followings are the list macros explained in TeX book p378
%
\newtoks\t@a \newtoks\t@b \newtoks\next@a
\newcount\num@i \newcount\num@j \newcount\num@k
\newcount\num@l \newcount\num@m \newcount\num@n
\long\def\l@append#1\to#2{\t@a={\\{#1}}\t@b=\expandafter{#2}%
                         \edef#2{\the\t@a\the\t@b}}
\long\def\r@append#1\to#2{\t@a={\\{#1}}\t@b=\expandafter{#2}%
                         \edef#2{\the\t@b\the\t@a}}
\def\l@op#1\to#2{\expandafter\l@opoff#1\l@opoff#1#2}
\long\def\l@opoff\\#1#2\l@opoff#3#4{\def#4{#1}\def#3{#2}}
%
% "sort@@" macros sort the given number-list and make reference-number
%
\newif\ifnum@loop \newif\ifnum@first \newif\ifnum@last
\def\sort@@#1{\num@firsttrue\num@lasttrue\sort@t#1}
\def\sort@t#1{\pop@@#1\to\num@i\rel@x%\message{numi**\the\num@i}%
            \ifnum\num@i=0 \num@lastfalse\let\next@a\rel@x%
            \else\num@looptrue%
                 \loop\pop@@#1\to\num@j\rel@x%\message{numj=\the\num@j}%
                    \ifnum\num@j=0 \num@loopfalse%
                    \else\ifnum\num@i>\num@j%
                         \num@k=\num@j\num@j=\num@i\num@i=\num@k%
                         \fi%
                    \fi%
                    \push@\num@j\to#1%
                  \ifnum@loop\repeat%
                  \let\next@a\sort@t%
            \fi%
%            \message{find>>>>\the\num@i}%
            \print@num%
            \next@a#1}
\def\print@num{%
              \ifnum@first%
                 \num@firstfalse\num@n=\num@i\number\num@i%
              \else%
                 \num@m=\num@i\advance\num@m by-\num@l%
                 \ifcase\num@m\message{%
                   *** WARNING *** Reference number %
                   [\the\num@i] appears twice or more!}%
                 \or\rel@x%
                 \else\num@m=\num@l\advance\num@m by-\num@n%
                    \ifcase\num@m\rel@x%
                    \or,\number\num@l%
                    \else-\number\num@l%
                    \fi%
                    \ifnum@last\num@n=\num@i,\number\num@i\fi%
                 \fi%
              \fi%
              \num@l=\num@i%
              }
\def\pop@@#1\to#2{\l@op#1\to\z@@#2=\z@@}
\def\push@#1\to#2{\edef\z@@{\the#1}\expandafter\r@append\z@@\to#2}

%%%%%%%%%%%%%%%%%%%%%%%%%%%%%%%%%%%%%%%%%%%%%%%%%%%%%%%%%%%%%%%%%%%%%
%
% Next the main part of keepref macros
%
% The idea of \append@cs macro is described
%                 in the TeXBook EXERCISE 7.10 (p41)
% The idea of \if@first@use macro is described
%                 in the TeXBook EXERCISE 7.7  (p40)

\def\append@cs#1=#2#3{\xdef#1{\csname%
                    \expandafter\g@bble\string#2#3\endcsname}}
\def\g@bble#1{}

\def\if@first@use#1{\expandafter\ifx\csname\expandafter%
                              \g@bble\string#1text\endcsname\relax}
\def\keep@ref#1#2{\def#1{0}\append@cs\y@@=#1{text}\expandafter\edef\y@@{#2}}
\def\keepref#1#2{\if@first@use#1\keep@ref#1{#2}%
                 \else\message{%
                    \string#1 is redefined by \string\keepref! %
                    The result will be .... what can I say!!}%
                 \fi}

\def\Null{0}
\def\get@ref#1#2{\def#2{\string#1text}}
\def\findref@f#1{%
                \ifx#1\Null \get@ref#1\text@cc\R@F#1{{\text@cc}}%
                \else\rel@x\fi}

\def\findref#1{\findref@f#1\ref@mark{#1}}

\def\find@rs#1{\ifx#1\endrefs \let\next=\rel@x%
              \else\findref@f#1\r@append#1\to\void@@%
                    \let\next=\find@rs \fi \next}
\def\findrefs#1\endrefs{\def\void@@{}%
                    \find@rs#1\endrefs\r@append{0}\to\void@@%
                    \ref@mark{{\sort@@\void@@}}}
\let\endrefs=\rel@x

%%%%%%%%%%%%%%%%%%%%%%%%%%%%%%%%%%%%%%%%%%%%%%%%%%%%%%%%%%%%%%%%%%%%%
%
%   Hereafter a simplified version of reference macros in PHYZNEW.tex
%
\newcount\referencecount     \referencecount=0
\newif\ifreferenceopen       \newwrite\referencewrite
\newdimen\refindent          \refindent=30pt
\def\refmark#1{\attach{\scriptscriptstyle  #1) }}
\def\NPrefmark#1{[#1]}
\def\ref@mark#1#2{\if#2,\rlap#2\refmark#1%
                  \else\if#2.\rlap#2\refmark#1%
                   \else\refmark{{#1}}#2\fi\fi}
\def\NPref@mark#1#2{\if#2,\NPrefmark#1,%
                  \else\if#2.\NPrefmark#1.%
                   \else\thinspace\NPrefmark{{#1}} #2\fi\fi}
\def\REF@NUM#1{\gl@bal\advance\referencecount by 1%
    \xdef#1{\the\referencecount}}
\def\R@F#1{\REF@NUM #1\R@F@WRITE}
\def\r@fitem#1{\par \hangafter=0 \hangindent=\refindent \Textindent{#1}}
\def\itemref#1{\r@fitem{#1.}}
\def\NPrefitem#1{\r@fitem{[#1]}}
\def\NPrefs{\let\refmark=\NPrefmark \let\itemref=\NPrefitem%
            \let\ref@mark=\NPref@mark}
\def\PRrefs{\let\refmark=\attach}
\def\R@F@WRITE#1{\ifreferenceopen\else\gl@bal\referenceopentrue%
     \immediate\openout\referencewrite=\jobname.ref%
     \toks@={\begingroup \refoutspecials}%
     \immediate\write\referencewrite{\the\toks@}\fi%
     \immediate\write\referencewrite{\noexpand\itemref%
                                    {\the\referencecount}}%
    \immediate\write\referencewrite#1}

\def\outrefs{\par\penalty-400\vskip\chapterskip
%   \spacecheck\referenceminspace
   \ifreferenceopen \toks0={\par\endgroup}%
   \immediate\write\referencewrite{\the\toks0}%
   \immediate\closeout\referencewrite%
   \referenceopenfalse \fi
%   \line{\fourteenrm\bf References\hfil}\vskip\headskip
   \centerline{\bf References}\vskip\headskip
   \input \jobname.ref
   }
\def\refoutspecials{\sfcode`\.=1000 \interlinepenalty=1000
         \rightskip=\z@ plus 1em minus \z@ }
%
% macros from "phyzzx"
%

\font\fourteenrm=cmr10 scaled\magstep2
\font\twelverm=cmr10 scaled\magstep1
\font\ninerm=cmr9	     \font\sixrm=cmr6
\newskip\chapterskip	     \chapterskip=\bigskipamount
\newskip\sectionskip	     \sectionskip=\medskipamount
\newskip\headskip	         \headskip=8pt plus 3pt minus 3pt
\newdimen\chapterminspace    \chapterminspace=15pc
\newdimen\sectionminspace    \sectionminspace=10pc
\newdimen\referenceminspace  \referenceminspace=25pc
\def\Textindent#1{\noindent\llap{#1\enspace}\ignorespaces}
\def\space@ver#1{\let\@sf=\empty \ifmmode #1\else \ifhmode
   \edef\@sf{\spacefactor=\the\spacefactor}\unskip${}#1$\relax\fi\fi}
\def\attach#1{\space@ver{\strut^{\mkern 2mu #1} }\@sf\ }
\def\spacecheck#1{\dimen@=\pagegoal\advance\dimen@ by -\pagetotal
   \ifdim\dimen@<#1 \ifdim\dimen@>0pt \vfil\break \fi\fi}
\catcode `\@=12
% ++++++++++++++++++++++++++++++++++++++++++++++++++++++++++++++++++++++++++
% ----- NOTE order of input FIGMACRO and keeprefMACRO -----
%\input figmacro
%\input refmacro
% ----- format ----- : use this format throughout
\magnification=\magstep1
\parindent=20 truept
\baselineskip 20 truept
\hsize=16.5 true cm
\vsize=23.6 true cm
% ----- format -----

\def \frac#1#2{ { #1 \over #2} }
\def \bra#1 {\langle #1 |}
\def \ket#1 {| #1 \rangle}
\def \braketa#1 {\langle #1 \rangle}
\def \braketb#1#2 {\langle #1 | #2 \rangle}
\def \braketc#1#2#3 {\langle #1 | #2 | #3 \rangle}
\def \gtsim{\mathrel{\hbox{\raise0.2ex
     \hbox{$>$}\kern-0.75em\raise-0.9ex\hbox{$\sim$}}}}
\def \ltsim{\mathrel{\hbox{\raise0.2ex
     \hbox{$<$}\kern-0.75em\raise-0.9ex\hbox{$\sim$}}}}

\def \dI { \Delta I }
\def \cJ { {\cal J} }
\def \cQ { {\cal Q} }

\def \rot { {\rm rot} }
\def \osc { {\rm osc} }
\def \eff { {\rm eff} }
\def \wob { {\rm wob} }
\def \NG { {\rm NG} }
\def \UR { {\rm UR} }
\def \PA { {\rm PA} }
\def \RPA { {\rm RPA} }

%
%  ********** Definition of math-boldface fonts family *********
%
\font\tenmib=cmmib10                     \skewchar\tenmib='177
\newfam\mibfam
\def\tenpoint{\relax
    \def\mib{\fam\mibfam \tenmib}
    \textfont\mibfam=\tenmib    \scriptfont\mibfam=\tenmib
    \scriptscriptfont\mibfam=\tenmib}
\tenpoint
\def\bfit#1{\mib #1}

%

% ----- references -----

\keepref\BMm{ \AA.~Bohr and B.~R.~Mottelson, {\it Nuclear Structure},
             Vol. I, (Benjamin, New York, 1969), Chap. 3-3b.}

\keepref\BMa{ \AA.~Bohr and B.~R.~Mottelson, {\it Nuclear Structure},
             Vol. II, (Benjamin, New York, 1975), Chap. 4, p.190 ff.}

\keepref\Mara{ E.~R.~Marshalek, Nucl. Phys. {\bf A331} (1979), 429.}

\keepref\Marb{ E.~R.~Marshalek, Phys. Rev. {\bf C11} (1975), 1426;
        Nucl. Phys. {\bf A266} (1976), 317.}

\keepref\Zel{ V.~G.~Zelevinsky, Nucl. Phys. {\bf A344} (1980), 109.}

\keepref\MJa{ I.~N.~Mikhailov and D. Janssen,
        Phys. Lett. {\bf B72} (1978), 303.}

\keepref\JMa{ D.~Janssen and I.~N.~Mikhailov,
        Nucl. Phys. {\bf A318} (1979), 390.}

\keepref\AKa{ C.~G.~Andersson et. al.,
        Nucl. Phys. {\bf A361} (1981), 147.}

\keepref\Ska{ J.~Skalski, Nucl. Phys. {\bf A473} (1987), 40.}

\keepref\Kura{ H.~Kurasawa, Prog. Theor. Phys. {\bf 64} (1980), 2055;
       {\bf 66} (1981), 1317;  {\bf 68} (1982), 1594.}

\keepref\SMa{ Y.~R.~Shimizu and K.~Matsuyanagi,
        Prog. Theor. Phys. {\bf 70} (1983), 144; {\bf 72} (1984), 799.}

\keepref\EMR{ J.~L.~Egido, H.~J.~Mang and P.~Ring,
        Nucl. Phys. {\bf A339} (1980), 390.}

\keepref\SKM{ Y.~R.~Shimizu, T.~Kisaka and M.~Matsuzaki,
           {\it Soryushiron Kenkyu} (Kyoto) {\bf 81} (1990), No.61, F182.}

\keepref\SMO{ Y.~R.~Shimizu and M.~Matsuzaki,
            in Proceedings of the International Conference on
       {\it Nuclear Structure at High Angular Momentum}, May 18-21, 1992,
        Ottawa, AECL-10613, pp.278-282.}

\keepref\SMB{ Y.~R.~Shimizu and M.~Matsuzaki,
            in Proceedings of the International Conference on
     {\it Physics from Large Gamma-ray Detector Arrays}, August 2-6, 1994,
        Berkeley, California, to be published.}

\keepref\MSM{ M.~Matsuzaki, Y.~R.~Shimizu and K. Matsuyanagi,
        Prog. Theor. Phys. {\bf 79} (1988), 836.}

\keepref\Mat{ M.~Matsuzaki, Nucl. Phys. {\bf A509} (1990), 269.}

\keepref\eER{ N.~R.~Johnson et. al., Phys. Rev. Lett. {\bf 40} (1978), 151.}

\keepref\eOS{ P.~Chowdhury et. al.,Nucl. Phys. {\bf A485} (1988), 136.}

\keepref\Fraa{ S.~Frauendorf, Nucl. Phys. {\bf A557} (1993), 259c.}

\keepref\Godm{ A.~L.~Goodman, Phys. Rev. {\bf C45} (1992), 1649.}

\keepref\Kane{ K.~Kaneko, Phys. Rev. {\bf 45} (1992), 2754;
                          Phys. Rev. {\bf 49} (1994), 3014.}

\keepref\Dn{ F.~D\"onau, Nucl. Phys. {\bf A471} (1987), 469.}

\keepref\HHa{ G.~B.~Hagemann and I.~Hamamoto,
               Phys. Rev. {\bf C40} (1989), 2862.}

\keepref\KOa{ A.~K.~Kerman and N.~Onishi,
          Nucl. Phys. {\bf A361} (1981), 179.}

\keepref\Oni{ N.~Onishi, Nucl. Phys. {\bf A456} (1986), 390.}

%\keepref\BesCS{ D.~R.~Bes, O.~Civitarese and H.~M.~Sofia,
%        Nucl. Phys. {\bf A370} (1981), 99.}
\keepref\BesT{ D.~R.~Bes and J.~Kurchan,
     {\it The Treatment of Collective Coordinates in Many-Body Systems},
     World Scientific Lecture Notes in Physics, Vol.34, 1990,
     World Scientific Publishing Co.}

\keepref\KBB{ J.~Kurchan, D.~R.~Bes and S.~Cruz Barrios,
        Phys. Rev. {\bf D38} (1988), 3309.}

\keepref\KBBt{ J.~Kurchan, D.~R.~Bes and S.~Cruz Barrios,
        Nucl. Phys. {\bf A509} (1990), 306.}

\keepref\SKa{ H.~Sakamoto and T.~Kishimoto,
        Nucl. Phys. {\bf A501} (1989), 205; 242.}

\keepref\Saka{ H.~Sakamoto, Nucl. Phys. {\bf A557} (1992), 583c.}

\keepref\BK{M.~Baranger and K.~Kumar, Nucl. Phys. {\bf A110} (1968), 490.}

\keepref\BMb{ \AA.~Bohr and B.~R.~Mottelson, {\it Nuclear Structure},
             Vol. II, (Benjamin, New York, 1975), Chap. 4, p.158 ff.}

\keepref\SRMP{ Y.~R.~Shimizu, J.~D.~Garrett, R.~A.~Broglia, M.~Gallardo,
               and E.~Vigezzi, Rev. Mod. Phys. {\bf 61} (1989), 131.}

\keepref\VHL{ M.~M.~Villard, Ph.~Hubert and R.~J.~Liotta,
              Physica Scripta, {\bf 26} (1982), 201.}

\keepref\LL{ L.~D.~Landau and E.~M.~Lifshitz, {\it Mechanics}, 3-rd ed.,
             (Pergamon, London, 1976), \S37.}

\keepref\LVH{ K.~E.~G.~L\"obner, M.~Vetter and V.~H\"onig,
              Nuclear Data Tables, {\bf A7} (1970), 495. }

% 164Dy JAERI data, Kusakari gr.
\keepref\KusaDy{ H.~Kusakari et. al.,
       {\bf JAERI Tandem Annual Report 1992}, 125.}

% 168Er Coul-ex. at LBL, Cline gr.
\keepref\KotEr{ B.~Kotli\'nski et. al., Nucl. Phys. {\bf A517} (1990), 365.}

% 162Dy BE2
\keepref\HelDy{ R.~G.~Helmer, Nuclear Data Sheets {\bf 64} (1991), 79.}

% 164Er BE2
\keepref\RonEr{ R.~M.~Ronningen et. al., Phys. Rev. {\bf C26} (1982), 97.}

% ------ refs by M.M. ---------

\keepref\mrefa{W.~Gast et al., Z.Phys. {\bf A318} (1984), 123.}

\keepref\mrefb{G.~Puddu, O.~Scholten and T. Otsuka,
                 Nucl. Phys. {\bf A348} (1980), 109.}

\keepref\mrefc{S.~T\"orm\"anen et al., Nucl. Phys. {\bf A572} (1994), 417.}

\keepref\mrefd{R.~Wyss et al., Nucl. Phys. {\bf A505} (1989), 337.}

\keepref\mrefe{D.~Ward et al., Nucl. Phys. {\bf A529} (1991), 315.}

\keepref\mreff{D.~C. Radford et al., Nucl. Phys. {\bf A545} (1992), 665.}

\keepref\mrefg{M.~Oshima et al., Phys. Rev. {\bf C37} (1988), 2578.}

\keepref\mrefh{I.~Hamamoto and B.~R.~Mottelson,
               Phys. Lett. {\bf B132} (1983), 7.}

\keepref\mrefi{A.~Ikeda, Nucl. Phys. {\bf A439} (1985), 317.}

\keepref\mrefj{N.~Onishi, I.~Hamamoto, S.~\AA berg and A.~Ikeda,
                  Nucl. Phys. {\bf A452} (1986), 71.}

\keepref\mrefk{M.~Matsuzaki, Nucl. Phys. {\bf A491} (1989), 433;
                         {\it ibid} {\bf A519} (1990), 548.}

\keepref\mrefl{R.~Bengtsson et al., Nucl. Phys. {\bf A415} (1985), 189.}

\keepref\mrefm{I.~Hamamoto and B.~Mottelson,
                Phys. Lett. {\bf B127} (1983), 281.}

\keepref\mrefn{A.~Ikeda and S.~\AA berg,
                Nucl. Phys. {\bf A480} (1988), 85.}

\keepref\mrefo{N.~Onishi and N.~Tajima,
                Prog. Theor. Phys. {\bf80} (1988), 130.}

\keepref\mrefp{M.~Matsuzaki, Nucl. Phys. {\bf A504} (1989), 456.}

\keepref\mrefq{C.~-H.~Yu et al., Nucl. Phys. {\bf A511} (1990), 157.}

\keepref\mrefr{M.~Matsuzaki, Phys. Rev. {\bf C46} (1992), 1548.}

% ------ start text -------

\centerline{ \bf Nuclear Wobbling Motion and Electromagnetic Transitions}

\vskip 4mm
\centerline{ Yoshifumi R. Shimizu and Masayuki Matsuzaki$^{*)}$ }
\vskip 2mm
\centerline{Department of Physics, Kyushu University 33, Fukuoka 812, Japan }
\centerline{$^{*)}$ Department of Physics, Fukuoka University of Education,}
\centerline{   Munakata, Fukuoka 811-41, Japan}

\baselineskip 15 truept
\vskip 5mm
\leftline{ \bf Abstract }
\vskip 2mm

  The nuclear wobbling motion is studied from a microscopic viewpoint.
It is shown that the expressions not only of the excitation energy
but also of the electromagnetic transition rate
in the microscopic RPA framework
can be cast into the corresponding forms of the macroscopic rotor model.
Criteria to identify the rotational band
associated with the wobbling motion are given,
based on which examples of realistic calculations are investigated
and some theoretical predictions are presented.

\baselineskip 22 truept
\vskip 5 mm
\leftline{ \bf \S1 Introduction }
%\vskip 2 mm

  The recent advent of new generation crystal ball detectors has been
opening a great possibility to explore a new area of the high-spin physics.
There are many interesting subjects which are waiting to be studied.
Among them, we would like to concentrate, in this paper, upon the nuclear
wobbling motion\findref\BMa,
 which is one of the "exotic" rotational motions in the
sense that the axis of rotation does not coincide with any of
the inertia axes of deformation.

  The nuclear wobbling motion has been considered by analogy with
the spinning motions of asymmetric top (classical rigid-body), where
perturbations are superimposed on the main rotation
around one of the principal axes with the largest moment of inertia.
When quantized, the energy spectra in the energy versus angular momentum plane,
which are nothing but those of the well-known
macroscopic triaxial rotor model (Davydov model),
are classified into two groups of rotational bands,
i.e., the "horizontal" and "vertical" sequences corresponding asymptotically
to the Regge trajectories associated with the largest and the smallest
moment of inertia.  In the high-spin limit\findref\BMa,
the physical meaning of those two sequences
appears to be more transparent by introducing an elementary excitation
of the "wobbling phonon" mode.  The horizontal sequences
parallel to the yrast line are rotational bands
in which zero, one, and two, etc, wobbling phonons are excited
in each intrinsic state,
while the vertical sequences starting from each yrast state
consist of the multiple wobbling phonon bands.
In fact, the $E2$ transition rates in the former ($|\Delta I| = 2$) are larger
than that in the latter ($|\Delta I| = 1$)
by an order of $1/I$ in the high-spin limit\findrefs\BMa\Mara\endrefs.

  It is very interesting to ask whether such an "exotic" rotational motion
is realized as a collective motion in atomic nuclei, because it directly
reflects the three-dimensional nature of rotational motions.
It should, however, be noticed that the concept of the three-dimensional
rotation is meaningful only when the motion of angular momentum vector
is traced in the "intrinsic" or the "body-fixed" frame,
where the macroscopic rotor model is formulated.
The very definition, however, of such a coordinate frame
is highly nontrivial from the microscopic point of view
in the general framework of the nuclear many-body
theory\findrefs\BesT\KOa\Oni\Kane\endrefs.

Detailed experimental investigation is only possible in
the discrete line spectroscopy so that the one or two phonons excited
horizontal sequences are the most promising targets for the study.
For such rotational bands near the yrast line,
the small amplitude approximation to the wobbling mode may be allowed.
Then the fully microscopic formulation is
possible\findrefs\Mara\MJa\JMa\Zel\KBBt\endrefs in terms of the random
phase approximation (RPA) on top of the cranked mean-field theory,
and the analogy to the macroscopic rotor model becomes
transparent.  Thus we mainly concentrate, in this paper,
on the yrast and the first excited wobbling bands
and investigate the possible consequences predicted
from the RPA theory\findref\Mara in realistic nuclei.
In the course of the investigation it will be shown that
not only the energy spectra but also the interband
($\Delta I = \pm 1$) electromagnetic transition probabilities
can also be expressed
in the same way as in the macroscopic rotor model in terms of
the microscopically defined "effective" moments of inertia\findref\Mara,
which are introduced for the wobbling eigen frequency.

  The paper is organized as follows: the $E2$ transitions in the RPA theory
is reviewed and is applied to the well-known $\gamma$-vibrational bands
in \S2.  Some basic ingredients in the macroscopic rotor model,
especially the expressions for the electromagnetic transition rates,
are reviewed in \S3 for the completeness.
The microscopic formalism is presented in \S4,
while some examples of realistic calculations are studied in \S5.
Main results are summarized in \S6.
Preliminary results of some part of the present work
has already been reported in Refs.\findrefs\SKM\SMO\SMB\endrefs.

\vskip 5 mm
\leftline{ \bf \S2 {\bfit E}2 Transitions for the $\gamma$-vibrational band}
%\vskip 2 mm

  The RPA theory on top of the cranked mean-field approach,
which is suitable for the high-spin states,  was first developed
in Ref.\findref\Marb and applied to the high-spin $\beta$- and $\gamma$-
vibrational bands\findrefs\EMR\SMa\endrefs.  The extension to the odd
nuclei, with special attention to how the electromagnetic
transition rates should be calculated, has been done in Ref.\findref\MSM.

  The RPA treatment of the vibrational excitations\findref\Marb
is based on the boson expansion theory; the lowest order vacuum states,
on which the RPA modes are excited, are described by the static
mean-field theory uniformly rotating around one of the inertia axes
(one dimensional cranking states).
Combined with the $1/I$ expansion technique,
the matrix element of the electromagnetic transition with multipolarity
$\lambda$ is expressed in the simple form;
$$
 {\cal M}(i \to f; \Delta I) \approx
   \braketc{f}{Q^{(E)}_{\lambda \mu=\Delta I}}{i} ,
               \eqno(2.1)
$$
where $\Delta I = I_f - I_i$ and the superscript $(E)$
means the electric part of the transition operator $Q_{\lambda \mu}$.
It should be stressed that the components of $Q^{(E)}_{\lambda \mu}$
are defined with respect to the rotation (cranking) axis.

  In the formula above only the lowest order in $1/I$ is retained.
Although we are mainly concerned with the high-spin limit
and consider only in the order of eq.(2.1) in the following sections,
it is worth while mentioning that the formula is applicable at low-spins
if the the geometry of the angular momentum vector
is properly taken into account.
A good example is the $M1$ transition in odd nuclei
at relatively low angular momenta\findref\Dn.
It is, however, noticed that the idea is more general
and is based on the observation that
the most important part of the spin-dependence,
which comes from the dynamics not from the kinematics of
the angular momentum algebra (the Clebsch-Gordan coefficients),
is contained in the right hand side through the spin-dependent
change of cranked wave functions.
In order to see this is really the case and to show the reliability
of the formula (2.1),
we compare, in the remaining part of this section, the results of
the RPA calculations for the low-spin $\gamma$-vibrational band\findref\SMa
in even-even nuclei with the experimental data.

  The quantities to be investigated are the $E2$ transitions
between the ground state band and the $\gamma$-vibrational band.
Therefore, the initial and the final state in eq.(2.1) are
$$
     \ket{i} = {\hat X}_\gamma^{(\pm) \dagger} \ket{\omega_\rot} ,\quad
     \ket{f} =  \ket{\omega_\rot} ,     \eqno(2.2)
$$
where $\ket{\omega_\rot} $ is cranked mean-field approximation
of the yrast states
with rotational frequency $\omega_\rot$,
and ${\hat X}_\gamma^{(\pm) \dagger}$ is the creation operator of
the RPA eigen mode at the corresponding frequency,
which smoothly continues to the $\gamma$-vibrational
solution at zero-frequency.  The superscript $(\pm)$\findref\SMa
denotes the signature quantum number carried by the RPA mode,
so that, for example, the $(+)$-band represents
the even-spin member and the $(-)$-band the odd-spin
if the vacuum $\ket{\omega_\rot} $ has zero signature.
There are five kinds of transitions associated with
$\Delta I = 0, \pm 1, \pm 2$, and the $B(E2)$-values of these transitions
are calculated by means of the electric (i.e. proton) part of
the operators\findref\MSM,
$$
    Q_{20} = \frac{1}{2} Q^{(+)}_0 + \frac{\sqrt{3}}{2} Q^{(+)}_2 ,
         \eqno(2.3a)
$$
$$
    Q_{2\pm 1} = \frac{i}{\sqrt{2}} \bigl( Q^{(-)}_1 \pm Q^{(-)}_2 \bigr) ,
         \eqno(2.3b)
$$
$$
    Q_{2\pm 2} = - \frac{1}{\sqrt{2}}
     \bigl( \frac{\sqrt{3}}{2} Q^{(+)}_0 + \frac{1}{2} Q^{(+)}_2
            \pm Q^{(+)}_1 \bigr) ,
         \eqno(2.3c)
$$
where $Q^{(\pm)}_K$ ($K=0,1,2$) are the signature coupled quadrupole
operators with the $z$-axis as the quantization axis,
while the rotation axis is chosen to be the $x$-axis as usual.

  Among the five transition amplitudes between the $\gamma$ and the ground
state band,
$$
     t[(Q^{(\pm)}_K)] \equiv \braketc{f}{Q^{(\pm)}_K}{i} =
  \braketc{\omega_\rot}
     {[Q^{(\pm)}_K , {\hat X}_\gamma^{(\pm) \dagger}]}{\omega_\rot}
     = \braketa{\, [Q^{(\pm)}_K , {\hat X}_\gamma^{(\pm) \dagger}] \,} ,
       \eqno(2.4)
$$
only the $K=2$ components are non-zero at $\omega_\rot=0$
from the selection rule
because the ground state under consideration is axially symmetric.
Moreover, the electric part of them are of the form
$$
    t[(Q^{(\pm)}_K)^{(E)}] = \cQ_\gamma \,
    ( \delta_{K2} + \delta_{K1} a_\gamma \omega_\rot )
      + O(\omega_\rot^2) ,
        \eqno(2.5)
$$
in the low frequency limit.  As it is mentioned we have to include
the effect of Clebsch-Gordan coefficients at the low-spin.  In this case
$\braketb{I_i 2 2 -2}{I_f 0} $ is necessary, which has the asymptotic
values in the large $I$ limit;
$$
     | \braketb{I_i 2 2 -2}{I_f 0} | \approx   \left\{
   \eqalign{ \sqrt{6}/4 & \quad \quad \Delta I =     0 , \cr
               1/2      & \quad \quad \Delta I = \pm 1 , \cr
               1/4      & \quad \quad \Delta I = \pm 2 . \cr }
       \right.       \eqno(2.6)
$$
Combining with eqs.(2.3) and (2.6) we have
the expression of the $B(E2) \equiv |{\cal M}(i \to f)|^2$,
which is valid in the low-spin limit,
$$
   \frac{B(E2)_{\gamma \to g}^{\Delta I}}{ \braketb{I_i 2 2 -2}{I_f 0} ^2 }
       = |\sqrt{2} \cQ_\gamma|^2
 \bigl( 1 + a_\gamma \Delta I \, \omega_\rot
      + O(\omega_\rot^2) \bigr)^2 ,       \eqno(2.7)
$$
where eq.(2.5) is inserted.  Using again the asymptotic relation
and the fact that the rotational frequency
can be related to the angular momentum
through the moment of inertia $\cJ$,
$$
    I_f(I_f+1) - I_i(I_i+1) \approx 2 I \, \Delta I ,\quad
           \hbar \omega_\rot \approx I/\cJ ,        \eqno(2.8)
$$
where $ I \equiv (I_f+I_i)/2 $, we finally find
$$
   \frac{\bigl[ B(E2)_{\gamma \to g}^{\Delta I} \bigr]^{\frac{1}{2}}}
           { \braketb{I_i 2 2 -2}{I_f 0} }
       \approx \cQ_t  \,
 \Bigl(1 + q \, [ I_f(I_f+1) - I_i(I_i+1) ] \Bigr) ,
            \eqno(2.9a)
$$
in the first order, which is nothing but the formula discussed
as the generalized intensity relation in Ref.\findref\BMb, where
$$
       \cQ_t \equiv | \sqrt{2} \cQ_\gamma | ,\quad
                 q \equiv a_\gamma /2 \hbar \cJ .
            \eqno(2.9b)
$$
It should be emphasized that the parameters
${\cal Q}_t$ and $q$ are now calculated microscopically
by means of the RPA theory at the finite rotational frequency.

  An example of the RPA amplitudes of the electric quadrupole transition
operators, eq.(2.4), as functions of the rotational
frequency is shown in Fig.1 for a typical well-deformed nuclei, $^{164}$Er.
The procedure of calculation is the same
as Refs.\findrefs\SMa\MSM\endrefs, except that the difference of
the oscillator frequency between neutrons and protons
in the Nilsson potential and of the oscillator length in the quadrupole
residual interaction\findrefs\BK\Saka\endrefs are properly
treated.\footnote{*)}{
  In Ref.\findref\SMa the RPA amplitudes of the $\gamma$-vibration
  for the {\it mass} quadrupole operators were shown.
  For the collective solutions the proper treatment of neutron and proton
  oscillator lengths makes the transition amplitudes
  of the electric operators approximately about $Z/A$
  of those of the mass operators,
  just as in the case of the static quadrupole moments,
  i.e. the mean values of the mass and electric quadrupole operators,
  see eq.(4.32) and discussions at the end of \S4.
}
Namely, the residual interactions of the monopole pairing and
the quadrupole interactions are used, the strengths of which are
determined so as to reproduce
the even-odd mass difference, and
the excitation energies of the $\beta$, $\gamma$ -vibrations ($K=0,2$)
and the zero-frequency (Nambu-Goldstone) mode ($K=1$),
at $\omega_\rot=0$, respectively.
After fixing the force strengths at $\omega_\rot=0$, the pairing
selfconsistent calculations are performed
as functions of the rotational frequency,
but the deformation parameters are fixed at the values
deduced from the experiments\findref\LVH, for simplicity.
Increasing the rotational frequency, the quasiparticle alignments
generally occur.  In order to identify individual rotational bands
with the same internal structure, are used the diabatic quasiparticle
orbits\findref\SMa, specifically the diabatic basis constructed
by the method of the $\omega_\rot$-expansion
up to the third order\findref\MSM.
The model space of the Nilsson orbits are chosen
as $N_\osc = 4-6$ for neutrons
and $=3-5$ for protons, which reproduces the transition amplitude
very well without using any kind of the effective charge (see below).

  {}From the figure it is seen that the amplitudes follow the low frequency
behaviours of eq.(2.5) and the parameters, ${\cal Q_\gamma}$ and $a_\gamma$,
are easily extracted.  The values thus obtained for some arbitrarily
chosen rare earth nuclei are summarized in Table 1,
where the parameters in eq.(2.9), $\cQ_t$ and $q$, are also included
and compared with the available experimental data.
Note that the moments of inertia for the ground state band
and the $\gamma$-vibrational band are the same in the low-spin limit
within the RPA theory.
It is not the case in experimental data, however.
This is because the difference of the moments of inertia results from the
higher order couplings between the vibrational and rotational motions,
which are not taken into account in the RPA.
Therefore two values of $q$ which are obtained by using two choices of
experimentally determined $\cJ = \cJ_g$
and $\cJ = \cJ_\gamma$ are included in the table.

  The absolute values of the amplitudes are well reproduced
in our calculations.
However, the agreement
might not be taken so seriously because the resultant RPA amplitude
depends on the size of the adopted model space.
It is known that the calculation with full
model space usually overestimates the experimental values
in the simple monopole pairing plus the quadrupole interaction model.
We will not discuss this point further in this paper.

  More important is that
not only the sign but also absolute values of the $q$ parameter
are well accounted for in the present RPA theory, which is calculated
from the ratio of the transition amplitudes and therefore more reliable
than the amplitudes themselves.
This clearly shows the rotation-induced change of the microscopic
structure of the vibrational motion, i.e. the $K$-mixing of the
transition amplitudes, see eq.(2.5), are correctly described
in the RPA theory.  In Fig.1 the appreciable deviations from the
lowest order relations, the order of $O(\omega_\rot ^2)$, are predicted
in the region of $\hbar \omega_\rot \gtsim 0.15$ MeV, for example,
the amount of the reduction of the $I_\gamma \to (I-1)_g$ transition rates
are larger than that of the $I_\gamma \to (I-2)_g$ transitions
(see eq.(2.3)).  We should, however, be careful to draw a definite conclusion:
at these moderate spins other higher order effects in $1/I$ neglected
in, e.g., eqs.(2.6), (2.8), are of the same order, and then
all the higher order terms should be consistently calculated,
which is out of the scope of the present calculations.

\vskip 5 mm
\leftline{ \bf \S3 Macroscopic Rotor Model }
%\vskip 2 mm

  The nuclear wobbling motion was originally predicted as a collective motion
in the macroscopic rotor model\findref\BMa.
In order to see the characteristic features and to compare with
the microscopic model, the energy spectra
and the $E2$ and $M1$ transition rates
in the high-spin limit are summarized in this section.

  The hamiltonian of the rotor model is written in terms of
the three moments of inertia around the principal axes,
$\cJ_x$, $\cJ_y$ and $\cJ_z$:
$$
      H_R = \frac{I_x^2}{2\cJ_x} + \frac{I_y^2}{2\cJ_y}
            + \frac{I_z^2}{2\cJ_z} ,        \eqno(3.1)
$$
where the angular momentum operators here are components with respect
to the {\it body-fixed frame}.
The energy spectrum of the above hamiltonian
is well known: it is specified by two quantum numbers
and they are given in the large $I$ limit explicitly by
$$
    E_R(I,n_w) = \frac{I(I+1)}{2\cJ_x} +
      \hbar \omega_w(I) \, \bigl( n_w + \frac{1}{2} \bigr) , \eqno(3.2)
$$
where $\cJ_x$ is assumed to be the largest and the main rotation
occurs around the $x$-axis.  The wobbling frequency is determined by
the well-known formula\findrefs\LL\BMa\endrefs,
$$
    \hbar \omega_w = I \, \sqrt{W_y W_z}=
     \hbar \omega_\rot \, \sqrt{\frac{(\cJ_x - \cJ_y)
           (\cJ_x - \cJ_z)}{\cJ_y \cJ_z}}, \eqno(3.3a)
$$
where
$$
     \hbar \omega_\rot \equiv I/\cJ_x ,   \eqno(3.3b)
$$
corresponds to the rotational
frequency around the main rotation axis and the quantities,
$$
    W_y \equiv 1/\cJ_z - 1/\cJ_x ,  \quad
    W_z \equiv 1/\cJ_y - 1/\cJ_x ,   \eqno(3.3c)
$$
are introduced.  The integer $n_w=0,1,2,...$ in eq.(3.2) is
the wobbling phonon number excited on the yrast states.
Then the horizontal and vertical sequences mentioned in \S1
are precisely
$$
    E^{\rm (hor)}_{|\Delta I| = 2}(I) = E_R(I,n_w), \quad
             n_w=0,1,2,...,   \eqno(3.4a)
$$
with $n_w$ specifying the yrast, yrare,... bands, and
$$
    E^{\rm (ver)}_{|\Delta I| = 1}(I) = E_R(I,n_w=I-K), \quad
             K=K_1,K_2,K_3,..., \eqno(3.4b)
$$
with $K$ specifying the band head spin.
Note that the number of phonons are changed in the vertical sequence,
while it is unchanged in the horizontal one.
These two classifications
correspond to the band structure connected by the $E2$ transitions
with $\Delta I = \pm 2 $ and $\Delta I = \pm 1$, respectively,
with the transition energies, in the lowest order in $1/I$,
$$
     E^{\rm (hor)}_\gamma = \mp 2 \hbar \omega_\rot , \quad
             (\Delta I = \pm 2) .      \eqno(3.5a)
$$
$$
     E^{\rm (ver)}_\gamma = \hbar \omega_w \mp \hbar \omega_\rot , \quad
             (\Delta I = \pm 1) .      \eqno(3.5b)
$$

  The $E2$ transition operator in the rotor model is derived by
the basic assumption that the quadrupole tensor of the rotor is
diagonal in the body-fixed frame; thus
$$
    Q_{2\mu}^{(E)} = e \Bigl( \frac{Z}{A} \Bigr) R^2
       \Bigr\{ \, \frac{\alpha_y + \alpha_z}{\sqrt{3}} D^{(2)}_{\mu,0}
  + \frac{\alpha_y - \alpha_z}{\sqrt{2}} (D^{(2)}_{\mu,2} + D^{(2)}_{\mu,-2})
        \Bigl\} ,   \eqno(3.6)
$$
and then the transition rates is given,
again within the large $I$ approximation, by
$$
    B(E2)^{\rm (hor)}_{\Delta I =\pm 2} \approx \Bigl( e \frac{Z}{A} \Bigr)^2
         \frac{1}{2} \, R^4 \, (\alpha_y - \alpha_z)^2 , \eqno(3.7a)
$$
$$
    B(E2)^{\rm (ver)}_{\Delta I =\pm 1} \approx \Bigl( e \frac{Z}{A} \Bigr)^2
         \frac{n_w}{I} \, R^4 \, \Bigl(
             \alpha_y \biggl(\frac{W_z}{W_y} \biggr)^\frac{1}{4}
         \mp \alpha_z \biggl(\frac{W_y}{W_z} \biggr)^\frac{1}{4}
                  \Bigr)^2 .        \eqno(3.7b)
$$
Here the deformation parameters $(\alpha_y, \alpha_z)$ are introduced
through the static quadrupole moments for later convenience;
$$
   \eqalign{
      R^2 \alpha_y \equiv &
           \frac{1}{2} \sqrt{\frac{15}{4\pi}}
               \int (x^2-z^2) \rho( {\bfit r} ) d{\bfit r} =
  \braketa{\, \frac{1}{2} Q^{(+)}_2 -\frac{\sqrt{3}}{2} Q^{(+)}_0 \,} , \cr
      R^2 \alpha_z \equiv &
           \frac{1}{2} \sqrt{\frac{15}{4\pi}}
               \int (x^2-y^2) \rho( {\bfit r} ) d{\bfit r} =
         \braketa{\, Q^{(+)}_2 \,} , \cr
         }                                     \eqno(3.8)
$$
in obvious notations.  These parameters,  $(\alpha_y, \alpha_z)$,
can be related to the usual "$(\epsilon_2,\gamma)$"-like
parametrization $(\epsilon_2^*, \gamma^*)$\footnote{*)}{
 These parameters coincide with the original
 ones only within the first order in $\epsilon_2$ because
 the original ones are defined with respect to the anisotropic
 harmonic oscillator frequencies.
 The Lund convention for the sign of triaxiality parameter $\gamma$
 is used throughout in this paper.}
which are defined by
$$
    \int x_k^2 \rho( {\bfit r} ) d{\bfit r} \equiv
     \frac{1}{3} R^2 \, \Bigl( 1+ \frac{4}{3} \epsilon_2^*
     \cos{(\gamma^* + \frac{2\pi}{3}k)} \Bigr) ,
        \quad (k=1,2,3 \equiv x,y,z)     \eqno(3.9)
$$
through
$$
  \alpha_y =
  \sqrt{\frac{5}{9\pi}} \epsilon_2^* \sin{(\gamma^*+\frac{4\pi}{3})},
     \quad
  \alpha_z =
 -\sqrt{\frac{5}{9\pi}} \epsilon_2^* \sin{\gamma^*} , \eqno(3.10)
$$
so that they represent the static triaxiality around
the corresponding $y,z$-axes.  In the same way ($\alpha_y - \alpha_z$)
is proportional to the static moments around the main rotation ($x$) axis
and then eq.(3.7a) simply denotes the usual stretched $E2$ transitions
with respect to the $x$-axis.
If we consider the lowest excited band, the 1-phonon wobbling band,
then $n_w=1$ in eq.(3.7b).

  In the same framework $M1$ transitions can be considered if an appropriate
magnetic property of the rotor is assumed.  According to Ref.\findref\VHL,
we assume that each component of the magnetic moment vector in the body-fixed
frame is proportional to that of the angular momentum,
$$
      m_i =  g_i I_i \quad (i=x,y,z) ,          \eqno(3.11)
$$
where $g_i$ $(i=x,y,z)$ denotes the g-factor with respect to
the $i$-th axis in the body-fixed frame.
Then the $M1$ transition operator is derived as\footnote{\dagger)}{
  The $M1$ operator used in Ref.\findref\VHL
  is different from eq.(3.12), which, we think, is incorrect.
  The resultant $B(M1)$ is also different. }
$$
   \eqalign{
    \mu_{1\mu} = \sqrt{\frac{3}{4\pi}} \, \mu_N
      \Bigl\{ g_x I^{(lab)}_\mu &
 + \frac{1}{2}(g_y+g_z-2g_x)(I_{+1}D^{(1)}_{\mu,+1}+I_{-1}D^{(1)}_{\mu,-1})
      \cr &
 - \frac{1}{2}(g_y-g_z)(I_{-1}D^{(1)}_{\mu,+1}+I_{+1}D^{(1)}_{\mu,-1})
                \Bigr\} } ,   \eqno(3.12)
$$
where $\mu_N$ is the nuclear magneton and
$I^{(lab)}_\mu$ means the angular momentum operator in the laboratory
frame so that the first term in the curly bracket does not contribute
to the transitions.  The $B(M1)$ is evaluated as in the same way as $B(E2)$;
$$
    B(M1)^{\rm (ver)}_{\Delta I =\pm 1}
      \approx \Bigl( \frac{3}{4\pi} \mu_N^2 \Bigr) n_w \frac{I}{4} \,
     \Bigl(  (g_y-g_x) \biggl(\frac{W_y}{W_z} \biggr)^\frac{1}{4}
         \mp (g_z-g_x) \biggl(\frac{W_z}{W_y} \biggr)^\frac{1}{4}
                  \Bigr)^2 .        \eqno(3.13)
$$

 Note that $B(E2)^{\rm (ver)}_{\Delta I =\pm 1} \propto 1/I$
and $B(M1)^{\rm (ver)}_{\Delta I =\pm 1} \propto I$ while
$B(E2)^{\rm (hor)}_{\Delta I =\pm 2}$ does not depend on $I$
(up to the leading order in $1/I$),
if all the parameters, the quadrupole deformations, $\alpha_y$ and $\alpha_z$,
the moments of inertia, $\cJ_x$, $\cJ_y$ and $\cJ_z$, and
the g-factors, $g_x$, $g_y$ and $g_z$, are constants against $I$.
Actually, all these parameters are not independent and, for example,
the quadrupole deformations completely determine the moments of inertia
in the original rigid-body model.
However, if one wants to apply the model to realistic nuclei,
one should consider that all the parameters depends smoothly
on the angular momentum,
and the moments of inertia do not necessarily take the rigid-body values.
As will be discussed in the following sections, the microscopically
calculated $\cJ_x$, $\cJ_y$ and $\cJ_z$ change
as functions of the rotational frequency and the relationship between
them are far from that of the rigid-body\findref\Mat.

  It is worth mentioning that the axially symmetric limit should be taken
with great care.  There are two kind of limits:
the "collective rotation" limit,
where $\gamma=0^\circ, -60^\circ$,
or equivalently either $\alpha_z=0$ or $\alpha_y=0$,
and the "non-collective rotation" limit
where $\gamma=60^\circ, -120^\circ$, or $\alpha_y=\alpha_z$.
In the former limit, it is easy to see that
there are no definite limiting expressions
for the wobbling energy and the $\Delta I = \pm 1$ $B(E2)$ and $B(M1)$,
as long as the detailed limiting behaviours
of three moment of inertia are not specified.
In contrast, one can always argue the latter limit, where, of course,
$\cJ_y=\cJ_z$ so that $W_y=W_z$.
Then, from eq.(3.7), the $\Delta I = +1$ vertical transition
as well as the $\Delta I = \pm 2$ horizontal transitions are prohibited.
In fact, in this axially symmetric limit of the "non-collective rotation",
the horizontal rotational sequences disappear and only the band heads
with $\braketa{I_x} =K$ are physically meaningful as the vacuum states
(not the vacuum band).
The collective rotation in this case, therefore,
is the rotation perpendicular to the symmetry axis
which is at the same time the non-collective rotation axis ($x$-axis),
and the band consists of multiple wobbling phonons with excitation energy
$$
     E^{\rm (ver)}_\gamma(K) = \hbar \omega_w(K) + \hbar \omega_\rot
             \approx  K/ \cJ_{\bot} , \quad
            (\cJ_{\bot} \equiv \cJ_y=\cJ_z) ,   \eqno(3.14)
$$
where eqs.(3.5b) and 3.3) are used,
and with the transfer angular momentum $\Delta I = +1$
excited on the high-$K$ band head (vacuum) states; namely,
$$
    E^{\rm (ver)}_{|\Delta I| = 1}(I) \approx
    \bigl( I(I+1)-K^2 \bigr) /2\cJ_{\bot} + {\rm const.} , \quad
        ( I \gtsim K),      \eqno(3.15)
$$
consistently to the large $I \approx K$ approximation.
Note that the rotational frequency $\hbar\omega_\rot = K/\cJ_x$
disappears in the final expressions as it should be because it is not
the collective rotational frequency in this case.
This kind of rotational band is specifically called as a "precession-band"
and often observed as a band excited on top of a high-$K$ isomer state
and has been well studied microscopically
in both realistic nuclei\findrefs\AKa\Ska\endrefs
and in a schematic model\findref\Kura.

  {}From the microscopic view point, there is a collective motion
even in the limit of axial symmetry of the "collective" rotation,
$\gamma=0^\circ$ and $-60^\circ$:
that is nothing but the $\gamma$-vibration with
signature $\alpha=1$ which has been just considered in the previous section.
The reason why the limit does not well-behave is
that the "body-fixed" frame or the "principal-axis" (see the next section)
frame of the quadrupole tensor is not well defined in this limit.
The calculation in the previous section
corresponds to the one in the "uniformly-rotating" frame,
which is always possible to define.
It will be shown, explicitly in the next section, that
the transformation to the "principal-axis" frame is impossible
in the "collective" rotation limit from the microscopic view point.

\vskip 5 mm
\leftline{ \bf \S4 Microscopic RPA Treatment }
%\vskip 2 mm

  Although the wobbling motion is a kind of oscillatory motion of
the rotation axis, the shape degrees of freedoms are necessary to consider.
Especially what is important is the non-diagonal parts of the quadrupole
tensor and thus we introduce,
$$
   \eqalign{
    Q_y \equiv Q_1^{(-)} = &
           - \frac{1}{2} \sqrt{\frac{15}{4\pi}}
               \sum_{a=1}^{A} (xz)_a   , \cr
    Q_z \equiv Q_2^{(-)} = &
           \,\, i \frac{1}{2} \sqrt{\frac{15}{4\pi}}
               \sum_{a=1}^{A} (xy)_a   , \cr
         }                    \eqno(4.1)
$$
which just appear as the $\Delta I = \pm 1$
$E2$ transition operators in eq.(2.3b).  Note that only the modes with
signature $\alpha=1$ are relevant in the RPA order
so that only these modes are considered in the following.
In higher order, however, one should also consider the remaining non-diagonal
component, $Q_x \equiv Q_1^{(+)}$\findrefs\KOa\Oni\endrefs, with all the RPA
eigen modes.

  In the RPA treatment, the in-band $\Delta I = \pm 2$ $E2$ transition
does not change the RPA phonon number and then the matrix elements
in the vacuum (or yrast) band and the one-phonon wobbling band are the same.
In contrast the interband transition from the one-phonon wobbling to
the vacuum band reflects the nature of the wobbling motion itself.
{}From the argument in \S2 (see eqs.(2.1) and (2.3c)) we have
$$
    B(E2)^{\rm (in-band)}_{\Delta I =\pm 2} \approx
         \braketa{\, Q_{2\pm 2}^{(E)} \,} ^2 =
              \Bigl( e \frac{Z}{A} \Bigr)^2
         \frac{1}{2} \, R^4 \, (\alpha_y - \alpha_z)^2 , \eqno(4.2a)
$$
where we have used eq.(3.8) with an understanding that the deformation
parameters are those associated with the vacuum states,
$\ket{\omega_\rot} $, and
$$
    B(E2)^{\rm (inter)}_{n,\Delta I =\pm 1} \approx \frac{1}{2}
     \Bigl( \bigl( \cQ_y(n) \mp \cQ_z(n) \bigr)^{(E)}
        \Bigr)^2,    \eqno(4.2b)
$$
where $n$ means that the $n$-th RPA eigen mode is considered and
the RPA amplitudes are defined as
$$
    \cQ_k(n) \equiv \braketc{n}{Q_k}{0} _\RPA =
            \braketa{ \, [X_n, Q_k] \, } \quad { (k=y,z) } ,   \eqno(4.3)
$$
with $X_n$ being the annihilation operator of the n-th RPA eigen mode.
Note that $Q_z$ is anti-hermite while $Q_y$ is hermite and this
is the reason why the sign in eq.(4.2b) is changed from that in eq.(2.3b).
Here and hereafter $\braketa{O} $ for any operator $O$ means the expectation
value with respect to the cranked state $\ket{\omega_\rot} $.
It is clear that the expression of the in-band $E2$ transition
formally coincides with those of
the horizontal transition in the macroscopic rotor model in eq.(3.7a).
How about the interband transition?
It is one of the main purpose of this section to clarify this point.

  The part of operators, $Q_y$ and $Q_z$, relevant in the RPA order
can be expanded in terms of the RPA eigen modes,
$$
   \eqalign{
      Q_y = &
      \sum_{n:all} \bigl( \cQ_y(n) X_n^\dagger + {\rm h.c.} \bigr) , \cr
      Q_z = &
      \sum_{n:all} \bigl( \cQ_z(n) X_n^\dagger - {\rm h.c.} \bigr) , \cr
      }          \eqno(4.4)
$$
Here $(n:all)$ means that the contribution of the Nambu-Goldstone (NG) mode,
$$
       X_\NG^\dagger = \frac{1}{\sqrt{2I}} ( J_z + i J_y ) ,
         \quad I \equiv \braketa{J_x} ,   \eqno(4.5)
$$
should be included:
$$
   \eqalign{
    \cQ_y(n=\NG)  = &
           - \frac{1}{\sqrt{2I}} \, 2R^2\alpha_y , \cr
    \cQ_z(n=\NG)  = &
          \, \frac{1}{\sqrt{2I}} \, 2R^2\alpha_z . \cr
         }                    \eqno(4.6)
$$
Namely the contribution of the NG mode corresponds to the "static"
deformation while those of the normal mode to the "dynamic" (or vibrational)
deformations.  It should be mentioned that there exists
a kind of "sum-rule"\footnote{*)}{
  Strictly speaking, it might not be called as a sum-rule because each term
  in the summation in eq.(4.7) is not positive definite.
}
which relates both contributions:
$$
    \sum_{n \ne \NG} \cQ_y(n) \cQ_z(n)  =
        \frac{R^4}{I} \, 2\alpha_y \alpha_z ,    \eqno(4.7)
$$
which can be easily verified from the identity,
$$
     [Q_y, Q_z] = 0  .   \eqno(4.8)
$$

  Although it is not stated explicitly we are working in the so-called
"uniformly rotating" (UR) frame in the sense that all observables
are based on the vacuum state $\ket{\omega_\rot} $
which is the cranked state
uniformly rotating around the one of the principal axes of the deformation
of the mean-field.  The wobbling motion naturally appears in the body-fixed
frame, or the "principal axis" (PA) frame\findref\Mara.
It is not a trivial matter to define the PA frame in the general
framework of the many-body problem.
One must introduce "gauge conditions"\findrefs\BesT\Kane\endrefs
which are common in the quantum theory with constraints.
According to Refs.\findrefs\Mara\KOa\endrefs, we impose the non-diagonal
part of the quadrupole tensor should vanish:
$$
     \bigl( Q_k \bigr)_\PA = 0 \quad (k=x,y,z) .        \eqno(4.9)
$$
The meaning of these condition are apparent.  The PA and UR frame picture
are related through the Euler angles which are now the dynamical variables,
for example,
$$
     \bigl( J_i \bigr)_\UR = \sum_{k=x,y,z}
    D_{ik}(\Theta) \bigl( J_k \bigr)_\PA ,        \eqno(4.10a)
$$
$$
     \bigl( Q_{ij} \bigr)_\UR = \sum_{k,l=x,y,z}
    D_{ik}(\Theta) D_{jl}(\Theta) \bigl( Q_{kl} \bigr)_\PA , \eqno(4.10b)
$$
where $Q_{ij}$ is the non spherically-coupled representation of the
quadrupole tensor, $Q_k=Q_{ij}$ (ijk-cyclic)
and $D_{ij}(\Theta)$ is the rank-1 $D$ function with
the additionally introduced Euler angles $\Theta$.
The three conditions, eq.(4.9), in principle,
define the microscopic structure of the three Euler angles in terms of
the complete set of the observables in the UR frame, where every microscopic
quantity is well-defined, and remove the redundancy between the
microscopic variables in the PA frame and the collective variables $\Theta$.
Of course it is not so simple: for instance, what is the ordering
between the Euler angle operators and the microscopic quadrupole tensor
in eq.(4.10b)?  One must work out very carefully
in a consistent framework\findref\BesT.

  Fortunately, it has been shown
that the small amplitude approximation makes the situation
quite simple\findrefs\Mara\Kane\endrefs.  The Euler angles can be
written in terms of the microscopic variables
and then the transformation from the UR to PA frame
can be done explicitly by using them within the RPA order\findref\Mara,
for example,
$$
   \eqalign{
     \bigl(   J_x \bigr)_\PA =&
            \bigl(   J_x \bigr)_\UR ,  \cr
     \bigl( i J_y \bigr)_\PA =&
       \bigl( i J_y - \frac{I}{2R^2\alpha_z} Q_z \bigr)_\UR , \cr
     \bigl(   J_z \bigr)_\PA =&
       \bigl(   J_z + \frac{I}{2R^2\alpha_y} Q_y \bigr)_\UR . \cr
       }      \eqno(4.11)
$$
Using the UR frame relations eq.(4.4), the angular momentum operators
in the PA frame are written in terms of the microscopic RPA eigen modes,
$$
   \eqalign{
     \bigl( i J_y \bigr)_\PA =&
       - I  \sum_{n \ne \NG}
     \Bigl( \frac{\cQ_z(n)}{2R^2\alpha_z} X_n^\dagger
             - {\rm h.c.} \Bigr) , \cr
     \bigl(   J_z \bigr)_\PA =&
      \,\, I  \sum_{n \ne \NG}
     \Bigl( \frac{\cQ_y(n)}{2R^2\alpha_y} X_n^\dagger
             + {\rm h.c.} \Bigr) , \cr
       }      \eqno(4.12)
$$
namely, the NG mode contribution disappears.
Apparently the transformation from the UR to PA frame
is possible only in the case when $\alpha_y \ne 0$ and $\alpha_z \ne 0$
(see eq.(4.21) below for more strict conditions).
Now it is easy to check, with the help of the sum-rule relation,
eq.(4.7), that (a part of) the commutation relation
in the PA frame, the sign of which is opposite to the one in the UR frame,
holds again within the RPA order\findref\Mara,
$$
   [\bigl( i J_y \bigr)_\PA, \bigl(   J_z \bigr)_\PA]
    = + I = \braketa{ \bigl(   J_x \bigr)_\PA } .  \eqno(4.13)
$$
It should be noticed that the transformation from the laboratory frame
to the UR frame is unitary while that from the UR to PA is not
as is clear from this commutation relation.
The physical reason why the transformation form the UR to PA is non-unitary
is apparent: the Euler angles in eq.(4.10) are not simple parameters
but are now considered to be the dynamical variables canonically
corresponding to the collective angular momenta,
which is a common feature to
the theory with constraints\findrefs\BesT\Kane\endrefs.

  Next, let us consider the dynamics in the PA frame.  The time-dependence
in the UR frame is governed by the hamiltonian,
$$
    H_\UR \equiv H - \omega_\rot J_x .       \eqno(4.14)
$$
Since the Euler angles are the dynamical variables describing the wobbling
of the rotation axis, the hamiltonian in the PA frame are modified,
$$
   \eqalign{
    H_\PA =&
       H - \Omega_x J_x - \Omega_y J_y - \Omega_z J_z \cr
           \approx &
       H_\UR - \Omega_y J_y - \Omega_z J_z \cr
      }    \eqno(4.15)
$$
where the operators in the right hand side are in the UR frame
(we omit $()_\UR$ hereafter)
 $\Omega_k \quad (k=x,y,z)$ are angular frequency operators
conjugate to the Euler angle operators and within the RPA order,
$$
     \Omega_x \approx \omega_\rot \gg \Omega_y , \, \Omega_z,  \eqno(4.16)
$$
reflecting the small amplitude RPA ansatz.
The microscopic structure of the angular frequency operators
are determined by the consistency conditions of
the gauge conditions\findrefs\Mara\Kane\endrefs,
$$
   i \frac{d}{dt} \bigl( Q_k \bigr)_\PA \equiv
     \bigl( [ Q_k, H_\PA] \bigr)_\PA = 0 \quad (k=x,y,z) .    \eqno(4.17)
$$
Again within the RPA order, we obtain
$$
   \eqalign{
    i\Omega_y =&
       - \frac{1}{2\hbar R^2\alpha_y} \bigl( [ H_\UR, Q_y] \bigr)_\PA
              =
       - \sum_{n \ne \NG} \Bigl\{
         \biggl( \omega_n \frac{\cQ_y(n)}{2R^2\alpha_y}
         + \omega_\rot \frac{\cQ_z(n)}{2R^2\alpha_z} \biggr) X_n^\dagger
               - {\rm h.c.} \Bigr\} , \cr
     \Omega_z =&
    \,\, \frac{1}{2\hbar R^2\alpha_z} \bigl( [ H_\UR, Q_z] \bigr)_\PA
              =
    \,\, \sum_{n \ne \NG} \Bigl\{
         \biggl( \omega_n \frac{ \cQ_z(n)}{2R^2\alpha_z}
         + \omega_\rot \frac{ \cQ_y(n)}{2R^2\alpha_y} \biggr) X_n^\dagger
               + {\rm h.c.} \Bigr\} , \cr
        }       \eqno(4.18)
$$
where the following results of the RPA eigen value problem in the UR frame
are used in the second equality in each equation,
$$
    H_\UR \approx \sum_{n:all} \hbar \omega_n X_n^\dagger X_n
          = \sum_{n \ne \NG} \hbar \omega_n X_n^\dagger X_n +
      \frac{\hbar \omega_\rot}{2I} \bigl\{ J_z^2 - (i J_y)^2 \bigr\} .
            \eqno(4.19)
$$
Eqs.(4.12) and (4.18) show that both the angular momentum vector and
the angular frequency vector precess around the main rotation axis (x-axis)
in the PA frame with the amplitudes,
$$
   \bigl(J_k \bigr)_\PA(n) \equiv
     \braketc{n}{\bigl(J_k \bigr)_\PA}{0} _\RPA , \quad
    \Omega_k(n) \equiv
     \braketc{n}{\Omega_k}{0} _\RPA  \quad  (k=y,z) .     \eqno(4.20)
$$
For this picture to be consistent the ratios of the dynamic to the static
deformation should be small, i.e.,
$$
  \eqalign{
      r_y(n) \equiv \cQ_y(n)/2R^2\alpha_y =&
          \, \bigl(J_z \bigr)_\PA(n)/I \sim O(1/\sqrt{I})  \ll 1 , \cr
      r_z(n) \equiv \cQ_z(n)/2R^2\alpha_z =&
           - \bigl(i J_y \bigr)_\PA(n)/I \sim O(1/\sqrt{I}) \ll 1 , \cr
     }    \eqno(4.21)
$$
Now the "effective" moments of inertia\findref\Mara for each RPA eigen mode,
which are naturally introduced through
$$
   \hbar \cJ_k^\eff(n) \equiv \bigl( J_k \bigr)_\PA(n) / \Omega_k(n)
            \quad  (k=y,z) ,     \eqno(4.22)
$$
are thus written as,
$$
   \eqalign{
     \hbar \cJ_y^\eff(n) =&
      \frac{I \, r_z(n)}{\omega_n r_y(n) + \omega_\rot r_z(n)} , \cr
     \hbar \cJ_z^\eff(n) =&
      \frac{I \, r_y(n)}{\omega_n r_z(n) + \omega_\rot r_y(n)} , \cr
          }  \eqno(4.23)
$$
where eqs.(4.12) and (4.18) and the definition (4.21) are used.
Introducing the quantities (c.f. eq.(3.3c)),
$$
   \eqalign{
     W_y(n) \equiv &
            \, 1/\cJ_z^\eff(n) - 1/\cJ_x , \cr
     W_z(n) \equiv &
            \, 1/\cJ_y^\eff(n) - 1/\cJ_x , \cr
      } \quad ( \cJ_x \equiv I/\hbar\omega_\rot )    \eqno(4.24)
$$
we have
$$
  \left\{  \eqalign{
    \hbar \omega_n \biggl(r_y(n)/r_z(n) \biggr) =& I \, W_z(n) , \cr
    \hbar \omega_n \biggl(r_z(n)/r_y(n) \biggr) =& I \, W_y(n) , \cr
    } \right.   \eqno(4.25)
$$
from which the well-known wobbling energy formula is obtained
as in the same way as in eq.(3.3a),
$$
    \hbar \omega_n = I \sqrt{W_y(n)W_z(n)} = \hbar \omega_\rot
     \sqrt{\frac{\bigl(\cJ_x - \cJ_y^\eff(n) \bigr)
                 \bigl(\cJ_x - \cJ_z^\eff(n) \bigr)}
           {\cJ_y^\eff(n)\cJ_z^\eff(n)} } .  \eqno(4.26)
$$
Note that the right hand side of this equation depends on the eigen mode
itself so that it is only a formal solution of the RPA equation,
to which eq.(4.25) is equivalent.
It may be worth noticing that eigen value eq.(4.25) can be
obtained by the Euler equation
for the PA angular momentum vector, which can be derived
from the TDHF variational principle\findrefs\Mara\KOa\endrefs.
In fact, the Euler equation with the PA hamiltonian eq.(4.15),
$$
    \frac{d}{dt} \bigl( J_i \bigr)_\PA =
  \sum_{jk}  \varepsilon_{ijk} \bigl( J_j \bigr)_\PA \Omega_k ,  \eqno(4.27)
$$
leads, for the component of the n-th RPA eigenmode,
$$
  \left\{ \eqalign{
   - \omega_n \bigl( i J_y \bigr)_\PA(n) =&
      \,\, \Omega_z(n) I - \omega_\rot \bigl( J_z \bigr)_\PA(n) , \cr
   - \omega_n \bigl(\, J_z \bigr)_\PA(n) =&
      \, i \Omega_y(n) I - \omega_\rot \bigl( i J_y \bigr)_\PA(n) , \cr
     } \right.         \eqno(4.28)
$$
which reduces to eq.(4.25) first by substituting
$\Omega_{y,z}(n)$ = $\bigl( J_{y,z} \bigr)_\PA(n)/\hbar \cJ_k^\eff(n)$
(eq.(4.22)) and next by
$\bigl(iJ_y \bigr)_\PA(n) = -I \, r_z(n)$
and $\bigl(J_z \bigr)_\PA(n) = I \, r_y(n)$ (eq.(4.21)).

  Since we are considering the case where eigen-energies of all RPA
solutions are real and positive,
the signs of $W_y(n)$ and $W_z(n)$ are the same and then the basic eq.(4.25)
can be formally solved for $r_y(n)$, $r_z(n)$,
$$
   \eqalign{
     r_y(n) =& \,\,\, c_n \frac{1}{\sqrt{2I}}
              \Bigl(\frac{W_z(n)}{W_y(n)} \Bigr)^\frac{1}{4} , \cr
     r_z(n) =& \,\sigma_n c_n \frac{1}{\sqrt{2I}}
              \Bigl(\frac{W_y(n)}{W_z(n)} \Bigr)^\frac{1}{4} , \cr
      }      \eqno(4.29)
$$
where $\sigma_n$ denotes the sign of $W_y(n)$ (= the sign of $W_z(n)$)
and the quantity $c_n$ is
the amplitude with which the n-th mode contributes to the sum-rule (4.7),
$$
  \sum_{n \ne \NG} r_y(n) r_z(n) = \frac{1}{2I} \iff
     \sum_{n \ne \NG} \sigma_n c_n^2 = 1 ,    \eqno(4.30)
$$
so that $c_n^2 \ll 1$ for non-collective solutions and $c_n^2 \approx 1$
for collective solutions.
Using the definition of $r_y(n)$ and $r_z(n)$ (eq.(4.21)), eq.(4.29) shows
that the transition amplitudes $\cQ_y(n)$ and $\cQ_z(n)$ can be
expressed in terms of the static deformation parameters and the effective
moments of inertia.  Inserting them into eq.(4.2b), thus,  we finally obtain,
$$
    B(E2)^{\rm (inter)}_{n,\Delta I =\pm 1} \approx
        \Bigl( e \frac{Z}{A} \Bigr)^2
         \frac{1}{I} \, R^4 \, c_n^2 \Bigl(
             \alpha_y \biggl(\frac{W_z(n)}{W_y(n)} \biggr)^\frac{1}{4}
      \mp \sigma_n \alpha_z \biggl(\frac{W_y(n)}{W_z(n)} \biggr)^\frac{1}{4}
                  \Bigr)^2 ,        \eqno(4.31)
$$
where we assumed
$$
     \cQ_k^{(E)}(n) \approx e \frac{Z}{A} \, \cQ_k(n)
               \quad (k=y,z) ,  \eqno(4.32)
$$
which are approximately satisfied for the (isoscalar) collective RPA
solutions.  Now the analogy with the eq.(3.7b) (with $n_w =1$)
in the macroscopic rotor model is clear: the main difference, except to the
assumption (4.32), is the fact that there are many solutions
in the microscopic RPA treatment and the amplitude $c_n^2 \ne 1$
and the sign $\sigma_n$ can be negative.
Note that the negative sign $\sigma_n=-1$ means
$\cJ_{y,z}^\eff(n) > \cJ_x$
or $\cJ_{y,z}^\eff(n) < 0$, which conflicts
our basic assumptions that the x-axis is the main rotation axis
and the introduced three moments of inertia are physically meaningful,
and therefore clearly shows that the interpretation of the n-th RPA mode
as a wobbling motion is not justified in such a case.
{}From this observation we use positiveness of the sign $\sigma_n$,
i.e. $ r_y(n) r_z(n) >0 $,
and $c_n^2 \approx 1$ as  criteria to identify the wobbling solutions
among the microscopic RPA eigen modes in later applications.

  For completeness, the microscopic expression for the $M1$ rate corresponding
to the $\Delta I = \pm 1$ $E2$ transitions, eq.(4.2b), is given:
$$
    B(M1)^{\rm (inter)}_{n,\Delta I =\pm 1} \approx \frac{1}{2}
       \Bigl( i \mu_y(n) \pm \mu_z(n) \Bigr)^2,    \eqno(4.33)
$$
with
$$
    \mu_k(n) \equiv \braketc{n}{\mu_k}{0} _\RPA =
     \braketa{ \, [X_n,\sqrt{\frac{3}{4\pi}}\mu_N ( g_l l_k + g_s s_k )] \, }
 \quad { (k=y,z) } ,   \eqno(4.34)
$$
where $g_l$ and $g_s$ are usual orbital and spin g-factors.
We here are mainly concerned with the even-even nuclei
so that only the vibrational contribution are included\findref\MSM.
In contrast to the case of $E2$ transitions, where is used
the same basic requirement as in the rotor model
such that the quadrupole tensor is diagonal in the PA frame,
there is no simple correspondence in $M1$ transitions
between the results of the microscopic RPA framework
and the macroscopic model.  This is because the basic assumption of
the $M1$ operator in the rotor model, eq.(3.11), is not always justified
from the microscopic viewpoint: for instance the effects of quasineutron or
quasiproton alignments on the $M1$ transitions cannot be taken into account.
If we nevertheless take the same assumption as the rotor model,
the operators responsible to the $M1$ transitions are\footnote{*)}{
  Note that only the component of the magnetic moment vector perpendicular
  to the angular momentum in the PA frame,
  ${\bfit m} - g_x {\bfit I}$, contributes to the transition.}
$i\mu_y \propto (g_y-g_x) (i J_y)_\PA$, $\mu_z \propto (g_z-g_x) (J_z)_\PA$.
Then using eqs.(4.11), (4.21) and (4.29) a similar expression to the one
in the rotor model, eq.(3.13), is easily obtained with the same notice
as in the $E2$ transitions, eq.(4.31).

%
% Discussions on collectivity and M1 for wobbling, by YRS
%
  In the following let us discuss general features
of the high-spin RPA eigen modes.
The properties of the solutions can be different from case to case
depending on the microscopic structure of vacuum states.
Until up to now, we have used the term "collective" somewhat ambiguously.
It might be necessary to make it more precise here.
The collectivity is usually referred, at least, in two different contexts:
(1) the transition matrix elements of characteristic observables to
the solutions, e.g. the $E2$ transition for the $\beta$, $\gamma$-vibrations
and the wobbling motion under discussions, are large, and
(2) the RPA amplitudes of phonon operators spread over
many two-quasiparticle states.
For example, it is the second one that is necessary
for eq.(4.32) to be valid.
Let us call the former the "large-transition" collectivity and the latter
the "spread-over" collectivity.
The shape vibrational motions in the ground state regions
almost always satisfy the conditions of both collectivities.
{}From our experiences, the spread-over collectivity
gets weaker in most cases for the RPA solutions
excited on the vacuum with
rotational aligned quasiparticles, e.g. the s-band, at higher spins.
One of the reasons for this is that the pairing correlations
are reduced by the blocking effects of the aligned particles\findref\SRMP.
In spite of the fact that the phonon amplitudes somewhat
tend to concentrate on few components, the large-transition collectivity
can survive even at higher-spins, since the energy lowering
of the two-quasiparticle states make the RPA eigen energy small
so that the backward amplitudes remain to be large.

  The discussions above indicates that the $M1$ amplitude
with using the microscopic $M1$ operator, eq.(4.34),
is small for the typical spread-over collective RPA solutions.
In fact a destructive interference occurs for the $M1$ operator,
because amplitudes of this type of solutions,
which are of isoscalar quadrupole character,
have many two-quasiparticle components
with phases favourable to the $E2$ operator.
Actually the quadrupole shape vibrations in
non-rotating nuclei have RPA amplitudes which transfer
the $K$ quantum number by even unit so that the transition
matrix elements for the $M1$ operator strictly vanish.
This is well established for the $\gamma$-vibrational bands
at low spins, where the $M1$ transitions are usually small.
Therefore the $\Delta I = \pm 1$ transitions are $E2$ dominant
in such cases.  It should, however, be noticed that
the spread-over collectivity of the RPA solutions are reduced
at higher-spins and then
the properties of individual two-quasiparticle states,
whose components are dominant in the RPA amplitudes,
manifest themselves as an enhancement of the $M1$ amplitudes.
We will see examples of such cases, where the $M1$ transitions
are non-negligible or even dominant, in the next section.

\vfill
\eject

\vskip 5 mm
\leftline{ \bf \S5 Examples of Numerical Calculations }
%\vskip 2 mm

  In the microscopic RPA formalism we have solutions of
vibrational character in many cases.
As it is clarified in the previous sections,
however, all of the solutions cannot necessarily be interpreted as
the wobbling-like motions even though they are collective enough
in the sense that the $E2$ transition amplitudes are large
(see discussions below eq.(4.31)).
Namely, in addition to the enough collectivity, the relative sign
of the transition amplitudes $\cQ_y(n)$ and $\cQ_z(n)$ should be
the same as that of the static asymmetries around $y$ and $z$-axes,
$\alpha_y$ and $\alpha_z$,
$$
    {\rm sign \,\, of} \,\, (\cQ_y(n)/\cQ_z(n)) =
    {\rm sign \,\, of} \,\, (\alpha_y/\alpha_z) .
                  \eqno(5.1)
$$
This relation was mentioned in another form in eq.(4.9) of Ref.\findref\Mat.
Only in such solutions the three moments
of inertia associated with the RPA solutions, eq.(4.22), are well defined
and are consistent to the picture of the wobbling motion.
Interestingly enough, the lowest collective RPA solutions, if they exist,
satisfy the condition in most of the cases we found.
It might be interesting to note that eq.(5.1) holds
also for $\gamma$-vibrational excitations in nuclei
with small selfconsistent $\gamma$ deformation:
It is a good approximation to set $\gamma=0$ in such nuclei
and then the same relative sign as the selfconsistent triaxiality
(eq.(5.1)) is obtained for the transition amplitudes
even with such an artificial setting of $\gamma=0$.

  More precisely speaking, one must also require the conditions eq.(4.21)
in order for the adopted small amplitude approximation to be valid.
This means that both static deformation around $y$ and $z$-axes
should be larger than the dynamic ones.
Therefore, one might expect the ideal wobbling motion
only in the region of equilibrium
shape with $\gamma \ne 0^\circ$ and $\gamma \ne -60^\circ$
in the $(\epsilon_2,\gamma)$ plane; i.e. an appreciable amount of
triaxiality is generally required.
Notice that the wobbling-like solutions can exist
in the axially symmetric cases with $\gamma = 60^\circ$ and $=-120^\circ$,
where $\alpha_y = \alpha_z < 0$ and $>0$, respectively,
and correspond to the $\Delta I =1$ excitations on a high-$K$ isomer states
(the precession-bands) as is discussed in the end of \S3.

  This apparent asymmetry between the four cases of
axially symmetric shapes are based on
the fact that the three rotation axes are not treated
equivalently: the $x$-axis is the main rotation axis
and the angular momenta  around the other two axes are small compared to it,
which is generally believed to be valid near the yrast states.
Such a treatment especially makes sense in the situation
where the quasiparticle alignments occur: then an appreciable part
of the main rotation are carried by single-particle degrees of freedoms,
while perpendicular components of the rotation
are only of collective nature.
Actually, we have found the wobbling-like
solutions only in the cases where aligned quasiparticles
exist in the vacuum state from which the RPA modes are excited,
which makes the vacuum time-reversal broken and induces
strong rotational $K$-mixing effects.

  With the general considerations above in mind we will show
some examples of the results of realistic calculations
and discuss the characteristic properties
of the obtained wobbling motions.

\vskip 3 mm
\leftline{ \it \S5-1 $\Delta I =\pm 1$ interband $E2$ transitions }
\vskip 2 mm

    Restricting ourselves to even-even nuclei,
the yrast band has naturally the quantum numbers,
$(\alpha,\pi)$ = (0,+) with even-spin values.  Then the (one-phonon)
wobbling band has $(\alpha,\pi)$ = (1,+) with odd-spin values.
Unfortunately, there are only few cases where $(\alpha,\pi)$ = (1,+) bands
are observed up to high-spin states.  Moreover it is generally
difficult to distinguish the collective phonon band from non-collective
quasiparticle bands, like the $AC$ or $BD$ two-neutron aligned bands
in the usual nomenclature of quasiparticle orbits,
only from the energy spectra.  Therefore, at least until now,
there is no definite evidence that the ideal
wobbling motions exist in atomic nuclei.

  One of the most important feature of the wobbling motions is that
the $E2$ amplitudes $\cQ_y(n)$ and $\cQ_z(n)$ have comparative magnitude,
in contrast to the case of the $\gamma$-vibrations in \S2,
where $|\cQ_y(=t[Q_{1}^{(-)}])| \ll |\cQ_z(=-t[Q_{2}^{(-)}])|$.
This property immediately leads that the one of the transitions with
either $\Delta I = +1$ or $= -1$ is much larger than the other
depending on the relative phase of $\cQ_y(n)$ and $\cQ_z(n)$,
see eq.(4.2b), or the triaxiality, see eq.(5.1).
This is situation quite analogous to the $M1$
transitions between the signature partner bands in odd nuclei,
where the single-particle matrix elements of the operators
$i\mu_y$ and $\mu_z$
between the signature partner bands play a similar role
to the two amplitudes $\cQ_y$ and $\cQ_z$, see Figure 2,
although the role played by the triaxiality is not necessarily the same.
For the ideal case of the wobbling motion where the formula of
the macroscopic model are valid, the staggering of
the $E2$ transitions between the $\Delta I = +1$ and $-1$ combined
with the excitation energy gives important information
on the triaxial deformation and/or the three moments of inertia;
for example,
$$
   \frac{\Delta B(E2)_{|\Delta I| = 1}}{B(E2)_{|\Delta I| = 2}}
   \equiv \frac{B(E2)_{\Delta I = -1} - B(E2)_{\Delta I = +1}}
             { B(E2)_{|\Delta I| = 2}}
     = \frac{8}{I}\frac{\alpha_y \alpha_z}{(\alpha_y - \alpha_z)^2},
              \eqno(5.2).
$$
just as in a similar kind of analysis for the staggering
of the $M1$ transitions in odd nuclei as in Ref.\findref\HHa.
The relation between the $B(E2)_{\Delta I = \pm 1}$ and
the triaxial equilibrium deformation is summarized in Figure 3.

  Although there is no definite evidence,
it is suggested in Ref.\findref\Mat
that the odd-spin sequence of the so-called
"extention of the $\gamma$-band"\findref\eOS
after the $g$-$s$ band crossing in $^{182}$Os
might be a candidate of the wobbling motion from the calculations
of the same RPA formalism.  It is instructive here to show
the results for this nucleus because the results of the calculations
show an ideal feature discussed in the previous sections.
This nucleus is supposed to be $\gamma$-soft and the two quasineutron
alignment induce the negative $\gamma$ deformation.
Although the potential energy surface are rather flat so that
it is difficult to determine the precise $\gamma$ value,
the RPA results with $\gamma=-18^\circ$ has been shown
to have desired property as an ideal wobbling motion\findrefs\Mat\SMO\endrefs.
It should be stressed that in order to obtain the wobbling-like RPA solution
the existence of the aligned quasiparticles in the vacuum configuration,
e.g. the $s$-band, is indispensable.

  The calculational procedures, the single particle potential,
the effective interactions and the model space etc,
are the same as those in \S2.
It should, however, be stressed that
the RPA dispersion equation with the NG modes being
explicitly decoupled\findref\SMa is used in this calculations.
For the case with $\gamma \ne 0^\circ$, then,
the quadrupole force parameters disappears for the signature
$\alpha=1$ sector.  Therefore the wobbling-like RPA solution are
obtained without any ambiguity depending on the choice of the force
parameters.  Just as in Ref.\findref\Mat,\footnote{*)}{
  In Refs.\findrefs\Mat\SMO\endrefs,
  the transition amplitudes are $10-30$\% larger because
  a larger model space ($N_\osc=5-7$ for neutrons and
  $=4-6$ for protons) than that adopted in this paper has been used.
  {}From the results in \S2 we hope the calculations in this paper
  are more realistic.
  In addition, the results in Ref.\findref\Mat were obtained
  by using the RPA equation without the NG-modes explicitly decoupled.
  The rotational frequency dependence of the excitation energy of the
  wobbling is thus a little bit different form that in Ref.\findref\SMO,
  where the same NG-decoupled equation as the present paper was used.
}
all the mean-field parameters are fixed, for simplicity,
with the values $\Delta_\nu=0.74$, $\Delta_\pi=0.98$ MeV and
$\epsilon_2=0.21$, $\gamma=-18^\circ$, which are roughly
appropriate for the $s$-band of $^{182}$Os.
The results of the RPA amplitudes for the lowest RPA solution
and the corresponding $B(E2)$ values are shown
as functions of the rotational frequency in Figure 4.
Since $-60^\circ < \gamma < 0^\circ$, $\alpha_y < 0$ and $\alpha_z > 0$,
and then the relative sign of $\cQ_y(n)$ and $\cQ_z(n)$
for the solution is negative (see eq.(5.1)).
Thus, the transitions from $(I-1)_\wob$ to $(I)_s$ is much stronger
than the one from $(I+1)_\wob$ to $(I)_s$,
as is clear in Fig.4 (see also Fig.3),
so that the $B(E2)$ values show remarkable zigzag behaviour.

  Unfortunately these transition rates have not yet been measured.
Since the $\gamma$-ray energy for the $(I-1)_\wob \to (I)_s$ transition
is much smaller, the $\gamma$ decay rate for this transition, which has
larger $B(E2)$, is hindered.  This might be the reason why it is difficult
to measure the transitions.
As clear from the general consideration, if the wobbling type
RPA solution exists for the positive $\gamma$ equilibrium shape,
the relative sign of $\cQ_y(n)$ and $\cQ_z(n)$ is positive and then
both the $B(E2)$ and the $\gamma$-ray energy
for the $(I+1)_\wob \to (I)_s$ transition is larger.
Therefore, it might be more easy to measure the transitions
in this case, although we could not find a good example of
such calculations.

  The $M1$ transitions between the wobbling and the $s$-band in this nucleus
is small, typically $B(M1)$($\mu_N^2$)/$B(E2)$(e$^2$b$^2$) $\approx$ 0.1,
except the highest frequency shown in Fig.4, where $M1$ amplitudes start
to grow rapidly.  The smallness of the $M1$ amplitudes comes from
the fact that the wobbling solution in this nucleus is an ideal
case and keep large collectivity up to rather high-spins
(see the discussion at the end of \S4),
although it reduces gradually as is seen from the $B(E2)$-values
in Fig.4.

  Finally, it should be mentioned that the second RPA solution
in this nucleus is also considerably collective\findref\Mat.
However, the relative sign of $\cQ_y(n)$ and $\cQ_z(n)$ is positive
and therefore it is a kind of vibrational mode
but is not of wobbling nature.

\vskip 3 mm
\leftline{ \it \S5-2 Effective moments of inertia }
\vskip 2 mm

  Another important outcome of the microscopic RPA formalism
is that the three effective moments of inertia\findref\Mara
can be calculated, which are highly nontrivial from the microscopic viewpoint.
It should be noticed that the ratios
$\cJ_y^\eff(n)/\cJ_x$ and $\cJ_z^\eff(n)/\cJ_x$
can be extracted from the ratios
$\omega_n/\omega_\rot$ and $\cQ_y(n)/\cQ_z(n)$
through eqs.(4.26) and (4.29);
the latter ratios are experimentally observable from the energy spectra
and the $B(E2)_{|\Delta I| =1}/B(E2)_{|\Delta I| =2}$.
It may therefore be interesting to see how the three moments of inertia
calculated from the wobbling RPA solution behaves as functions of
deformation parameters, especially the triaxiality $\gamma$.
Their dependence on the rotational frequency have been
already studied in Ref.\findref\Mat for $^{182}$Os, and
will be presented for $^{124}$Xe in the next subsection.
In the case of the quadrupole residual interaction,
a simple formula for $\cJ^\eff_{y,z}(n)$ in terms of the 2$\times$2-coupled
RPA dispersion equations exists\findrefs\SMa\Mat\endrefs.
We have used this formula in the following calculations
in place of the original definition, eq.(4.22).

  Of course the true equilibrium shape has a definite
$\gamma$ deformation.  Here, however, we fixed the other parameters
of the mean-field potential and have performed the RPA calculations
with changing the $\gamma$ deformation as a free parameter.
Thus, it should be considered to be a kind of theoretical simulations
how the nucleus behaves if the triaxiality is artificially changed.
The results for the three sectors in the $(\epsilon_2,\gamma)$ plane,
$-120^\circ < \gamma < -60^\circ$, $-60^\circ < \gamma < 0^\circ$ and
$0^\circ < \gamma < 60^\circ$, in each of which a representative nucleus,
$^{176}$Hf, $^{182}$Os and $^{148}$Gd, is respectively selected,
are shown together in one panel in Figure 5.
Note that the yrast sequences of $^{176}$Hf and $^{148}$Gd
are known to be composed of non-collective rotations
with prolate and oblate shapes,
i.e. $\gamma=-120^\circ$ and $=60^\circ$, respectively.
Since all the other mean-field parameters are fixed with
neglecting the selfconsistency, the wobbling-like solutions
are not obtained for all values of $\gamma$ deformation,
and, moreover, the solutions can be discontinuous as functions of $\gamma$
because of the existence of virtual level crossings of the quasiparticle
orbits when changing $\gamma$.

  In spite of the deficiencies of
this relatively simple-minded calculations,
it is instructive to see the {\it microscopically derived}
$\gamma$ dependence of the three moments of inertia, which
is neither of irrotational like nor of rigid-body like.
In these calculations all the vacuum states,
on which the RPA mode excited,
have aligned quasiparticles, which is essential to obtain the
wobbling-like solutions as is mentioned above,
so that the $\cJ_x=I/\hbar \omega_\rot$ (kinematical moment of inertia)
changes very gradually and never vanish.
Thus the $\gamma$ dependence of the three moments
cannot be that of irrotaional.
On the other hand $\cJ_k^\eff \quad (k=y,z)$
are dynamical moments of inertia and take as small values as zero.
Thus the $\gamma$ dependence of the three moments
cannot be that of rigid-body, either.

  It is very interesting to see how the three moments of inertia,
which can be experimentally extracted from the observed wobbling motions,
if they exist, behave.  For this purpose,
the triaxial deformation should
be also determined from the independent experimental observables,
e.g. by the $M1$ transitions in neighbouring nuclei.

\vskip 3 mm
\leftline{ \it \S5-3 Wobbling motion in $^{124}$Xe }
\vskip 2 mm

 As pointed out above, one of possible candidates of wobbling motion
might be the high-spin continuation of the odd-spin members of
the so-called $\gamma$-band.
Actually a scenario of character change from the $\gamma$-vibrational
to the wobbling-like band was first pointed out in Ref.\findref\MJa,
and such a trend, though not well developed because of the small triaxiality,
are suggested for $^{164}$Er in Ref.\findref\SMa.
The essential point is that a structural change of the vacuum caused by
the quasiparticle alignment transfers an appreciable amount of $K=1$
quadrupole strength from the Nambu-Goldstone mode to the normal modes.
Although it is difficult to predict precisely
in which nuclei wobbling mode appears
because this mode is an outcome of subtle
interplay between rotation, quadrupole and pairing correlations in realistic
nuclei, we have at least up to now two typical candidates of wobbling motion
in observed bands with negative $\gamma$ deformation:
One is in $^{182}$Os after the $(\nu i_{13/2})^2$ alignment discussed in
Ref.\findref\Mat and above, and the other is in $^{124}$Xe \findref\mrefa
after the $(\nu h_{11/2})^2$ alignment being discussed in the following.
   \footnote{*)}{We thank R. Wyss for informing us of these data.}
An important characteristic common to both nuclei is that the low-spin
spectrum is almost rotational but relatively $\gamma$-soft while
negative $\gamma$ deformation,
$-30^\circ < \gamma < 0^\circ$ in $^{182}$Os and
$-60^\circ < \gamma < -30^\circ$ in $^{124}$Xe,
is stabilized above the first band crossing
caused by the high-$\Omega$ quasiparticles.
Here we note that in $^{182}$Os the $\gamma$-ray
which links (the candidate of) the odd-spin members
of the $\gamma$ band above and below the band crossing
has not been observed.
{}From this fact, we imagine
it might well be the case that there might exist
wobbling-like bands but have been assigned incorrectly
as non-collective two-quasiparticle bands in other nuclei
because of the lack of the linking transitions.
A possible mechanism of this lack will be discussed later
in \S5-4 for the case of $^{126}$Ba.

 The low-spin nuclear structure of Xe isotopes has been studied extensively by
means of the interacting boson model\findref\mrefb
and known to depend smoothly on
the neutron number.
A recent study\findref\mrefc, however, pointed out that above the first band
crossing there are some properties which show sudden changes between
$A=120$ and 122 due to the shape coexistence; the lighter isotopes remain
nearly axially symmetric while the heavier ones become triaxial
with $\gamma < 0$.
In particular, in $^{120}$Xe,
a candidate of the even-spin continuation of the $\gamma$ band above
the {\it second} band crossing has been found for the first time,
whereas in $^{124}$Xe,
the odd-spin sequence has been known to extend up to
higher-spins than the even-spin one \findref\mrefa
and it shows well-developed wobbling character.
We mainly concentrate on $^{124}$Xe in the following,
although similar results can be obtained for other isotopes
with similar triaxial deformation.

 The result of the RPA calculation for $^{124}$Xe based on the
$(\nu h_{11/2})^2$ aligned configuration is shown in Figure 6 %\mfiga
together with the experimental data seen from
a reference which makes the routhian of the $s$-band flat.
The method of the calculation are the same as in the previous sections
but the smaller model space, $N_\osc=3-5$ for both neutrons and protons
are used in this region of nuclei.
This calculation have been done with using the deformation parameters
$\epsilon_2=0.19$ and $\gamma=-45^\circ$, which are
taken from the Total Routhian Surface (TRS) calculation
at $\hbar\omega_\rot= 0.3$ MeV in Ref.\findref\mrefd.
The hexadecapole deformation is neglected for simplicity.
Although the meaning of the shape parameters
is slightly different because of the different single-particle potential
adopted in Ref.\findref\mrefd,
we confirmed the stability of results of our calculations
against possible small change of shape parameters.
The pairing gaps $\Delta_\pi=$1.1 MeV and $\Delta_\nu=$0.9 MeV
typical for the low spin part of the neutron $s$-bands
in this region are used.
This RPA calculation reproduces the data very well.
Note, again, that we do not have any adjustable parameter
in the step of RPA for the signature $\alpha=1$ sector.
Calculated effective moments of inertia are shown in Figure 7. %\mfigb
They vary as functions of the rotational frequency in spite of
the fact that the nuclear shape is fixed
as in the case of $^{182}$Os\findref\Mat,
but the dependence is much weaker in $^{124}$Xe.

  Calculated $E2$ and $M1$ transition amplitudes are shown
in Figures 8 and 9, respectively, as functions of the rotational frequency,
where the standard values of $g_l$ and $g_s^{\rm(free)}$\findref\BMm,
and $g_s^{\rm(eff)}/g_s^{\rm(free)}=0.7$ are used in this and the
following calculations for the $M1$ operators, eq.(4.34).
In the figures, the transition probabilities are also included.
Since the selfconsistent $\gamma = -45^\circ$, the relative phase
between $\cQ_y$ and $\cQ_z$ is negative so that the $\Delta I = +1$
transitions are stronger as in the same way as in $^{182}$Os.
It is, however, noticed that the collectivity of the RPA solution,
especially the spread-over one, is weaker than that in $^{182}$Os,
and consequently the $M1$ transitions is stronger in $^{124}$Xe.
Actually, as is seen from Figs.8 and 9, $M1$ transitions can compete
with $E2$ depending on the transition energies.
%Although we do not have any definite reason yet, a phase rule
%for the $M1$ amplitudes,
%$$
%    {\rm sign \,\, of} \,\, (i\mu_y(n)/\mu_z(n)) =
%    - {\rm sign \,\, of} \,\, (\alpha_y/\alpha_z) ,  \eqno(5.3)
%$$
%which is very similar to that for the $E2$, eq.(5.1),
%holds in almost all the cases we have examined.
%Therefore the zigzag behaviour of the $E2$ and $M1$ transitions
The zigzag behaviour of the $E2$ and $M1$ transitions
are the same in this case, see Figs.8 and 9.
It might be interesting to point out that trends
that the $E2$ transitions decrease against
the spin while the $M1$ transitions increase coincide with
the prediction of the macroscopic rotor model in this case,
see eqs.(3.7b) and (3.13).

   We use the quadrupole residual interaction which has two components,
$K=1$ and $K=2$, in the signature $\alpha=1$ sector.  Note that
a triaxial deformation mixes $K$ quantum number by two unit so that
the $K=1$ and 2 components of the interaction couple
only through the rotational motion.
It is the interplay of these two components intermediated by
the rotational coupling that brings about the character change of the
lowest collective excitation.
To look at this mechanism more closely,
it is useful to see the second-lowest RPA
solution together with the lowest one that we have been discussing.
The excitation energies of the two solutions
and the $K=1$ and 2 quadrupole transition amplitudes,
$\cQ_y(=t[Q_{1}^{(-)}])$ and $\cQ_z(=-t[Q_{2}^{(-)}])$,
associated with them, calculated at $\hbar\omega_\rot = 0.3$ MeV
are shown in Figure 10 %\mfigc
as functions of $\gamma$ from $-60^\circ$ to $0^\circ$.
The calculation have been done using the same
mean-field parameters (except $\gamma$) as in Fig.6. %\mfigb
The second-lowest solution
shares an appreciable amount of $\cQ_y(n)$ and $\cQ_z(n)$ strength
with the lowest one but with the different relative sign between them.
The similar situation has already been encountered
in the case of $^{182}$Os \findref\Mat.
In order to interpret these results, here we consider
the axially symmetric limits, $\gamma=0^\circ$ and $-60^\circ$,
where the symmetry axes are $z$ and $y$ axis, respectively.
Since the role of $y$ and $z$ axes are interchanged
(with the rotation axis unchanged) in both cases,
the modes with $|\cQ_z| \gg \,(\ll)\, |\cQ_y|$ in the $\gamma=0^\circ$
on one hand corresponds to
those with $|\cQ_y| \gg \,(\ll)\, |\cQ_z|$ in $\gamma=-60^\circ$
on the other hand.  We can see clearly in Fig.10
that this correspondence is actually holds and, moreover,
that the strongest $K$-mixing of the two amplitudes occurs
at around $\gamma=-30^\circ$.
If we call the mode with $|\cQ_z| \gg |\cQ_y|$ in the $\gamma=0^\circ$
"$\gamma$-vibration like",
then so is the second-lowest solution in $^{124}$Xe
at the rotational frequency shown.  In contrast the lowest solution
is "$\gamma$-vibration like"
in $^{182}$Os at $\hbar\omega_\rot \ltsim 0.25$ MeV
as is shown in Fig.4.  It should, however, be noticed that
the lowest solution follows the "phase rule", eq.(5.1), which
allows us to interpret the lowest solution
as a wobbling motion in both cases.
According to the previous discussions
the $\Delta I = +1$ $E2$ transitions are larger
for the lowest solution because $(\cQ_y/\cQ_z)< 0$,
while the $\Delta I = -1$ $E2$ transitions are larger
for the second-lowest solution.
No candidates of the second-lowest collective excitation
has been observed so far.
According to the TRS calculation\findref\mrefd, the equilibrium shape of the
$(\nu h_{11/2})^2$ band of $^{122}$Xe is $\gamma\simeq-30^\circ$, where the
excitation energy of the second-lowest solution becomes low and the difference
between $B(E2)_{\Delta I = \pm 1}$ becomes conspicuous due to the $K$-mixing.
So we think $^{122}$Xe is a more promising nucleus for which our theoretical
prediction can be tested.

\vskip 3 mm
\leftline{ \it \S5-4 Continuation of $\gamma$-band in $^{126}$Ba }
\vskip 2 mm

   Next we study the high-spin continuation of the $\gamma$ band
in $^{126}$Ba\findref\mrefe.
This nucleus is also relatively $\gamma$-soft at low spins, while the first
band crossing is caused by the low-$\Omega$ $(\pi h_{11/2})^2$ in contrast to
the isotone $^{124}$Xe studied above.
See Fig.6 of Ref.\findref\mrefd for a summary of the data.
The alignment of high-$j$, low-$\Omega$ quasiparticles drives the nucleus
towards $\gamma \gtsim 0$.
We have performed the RPA calculation for excitations on top of the
$(\pi h_{11/2})^2$ configuration
adopting $\epsilon_2 =0.24$ and $\gamma=0$ according to
the TRS calculation\findref\mrefe just as in the case of $^{124}$Xe.
On the other hand the pairing strengths are chosen
so as to reproduce their band crossing frequencies
(Fig.12 of Ref.\findref\mrefe) both for proton
($\hbar\omega_{\rm c}\simeq 0.35$ MeV) and neutron ($\simeq 0.44$ MeV)
when the gap selfconsistent calculation has been done.
We also confirmed that the parameters thus adopted reproduce the observed
$B(E2:2^+_1\rightarrow0^+_1) = 1.9 \pm 0.2 {\rm e}^2{\rm b}^2$
at the ground state.
The calculated pairing gaps for the proton $s$-band are
$1.16\geq\Delta_\nu \geq 1.03$ MeV and $0.76\geq\Delta_\pi \geq 0.71$ MeV
for $0.4\leq\hbar\omega_\rot\leq0.5$ MeV.
Since $\gamma \approx 0^\circ$ in this case we cannot use the RPA equation
with the NG-mode fully decoupled.  We therefore have used the equation
with the NG-mode partially decoupled\findref\SMa suitable for $\gamma=0^\circ$,
where the force parameter $\chi_2^{(-)}$ remains
and has been fixed so as to give the correct excitation energy of
the $\gamma$ vibration on the ground state as
$\hbar \omega_\gamma(\hbar \omega_\rot=0) = 0.873$ MeV.

  The result of the calculated routhian is shown in Figure 11 %\mfigf
together with the data seen from a reference
which makes the routhian of the $s$-band flat,
while the $E2$ and $M1$ amplitudes are depicted
in Figures 12 and 13 with corresponding transition probabilities included.
The excitation energy decreases as the rotational frequency increases both in
the data and calculation; this is contrary to the macroscopic wobbling motion.
This result, together with the fact that the calculated
signature-splitting of energy between $\alpha=0$ (not shown)
and $=1$ RPA solutions is very small, favours an interpretation
that the band under consideration remains to be $\gamma$-vibration like
rather than evolving to wobbling like above the band crossing,
as in the case of $^{164}$Er\findref\SMa.
In fact, the mixing of the $K=1$ component is not so strong,
as it is shown in Figure 12, which supports this interpretation.
The collectivity measured by the sum of the squared backward amplitudes
is similar to the $^{124}$Xe case; $\sim 0.6 $ in both nuclei.
Consequently the magnitudes of the interband $B(E2)$ are also similar,
although the $\Delta I = -1$ $B(E2)$ is now
larger than the $\Delta I = +1$ one
because $\cQ_y/\cQ_z >0$, which is characteristic to
small positive-$\gamma$ nuclei, see eq.(5.1).
This trend of the triaxiality reflects the shape driving effect
of the low-$\Omega$ aligned $(\pi h_{11/2})^2$.
%As it is pointed out in eq. (5.3),
%the zigzag behaviour of the $M1$ transitions
%is the same as that of the $E2$ transitions,
%similarly to the case of $^{124}$Xe.
The zigzag behaviour of the $M1$ transitions are not pronounced at all
in this case.  This is because the $\mu_z$ amplitudes are very small.
In addition, $M1$ transitions is decreasing as a function
of the rotational frequency,
which is in contradiction to the prediction of the macroscopic formula.

  An interesting feature of the data in this $^{126}$Ba nucleus
is that two branching ratios for decays
from the members of the band under consideration,
$T(17_{s\gamma}^+\to 16_s^+)/T(17_{s\gamma}^+\to 15_{s\gamma}^+)$ and
$T(19_{s\gamma}^+\to 18_s^+)/T(19_{s\gamma}^+\to 17_{s\gamma}^+)$,
where $s$ and $s\gamma$ stand for the $s$-band and
the $\alpha=1$ $\gamma$-band excited on top of the $s$-band, respectively,
were measured.  Using the calculated values
$B(E2:I_{s\gamma}\to (I-1)_s)=0.044$ ${\rm e}^2{\rm b}^2=12$ W.u.
(c.f. Fig.12) and $B(E2:I_{s\gamma}\to (I-2)_{s\gamma}) \simeq
 B(E2:(I-1)_s\to (I-3)_s)=0.69{\rm e}^2{\rm b}^2 = 184$ W.u.
at $\hbar\omega_\rot = 0.4$ MeV
and measured $\gamma$ ray energies, the estimated branching ratio is
$T(E2:I_{s\gamma}\to (I-1)_s)/T(E2:I_{s\gamma}\to (I-2)_{s\gamma}) \sim 0.05$,
which is an order of magnitude smaller than the measured one\findref\mrefe.
This apparent contradiction between the data and the calculation
can be solved by the specific property of the RPA solution in this case
such that the $M1$ transition is very strong, as is shown in Fig.13.
Compare the $B(M1)$ values with those of $^{124}$Xe in Fig.9.

  As is pointed above the collectivity of the calculated solution
in this case is more or less similar to that in $^{124}$Xe, which
is a little bit weaker than that in $^{182}$Os.
Actually, looking into the details of the microscopic structure of the RPA
phonon, we have found appreciable amount of concentrations of
the phonon amplitudes to the two-quasiparticle components with
$(\pi h_{11/2})^2$ in $^{126}$Ba and with $(\nu h_{11/2})^2$ in $^{124}$Xe,
in contrast to the strong spread-over collectivity seen
in the case of $^{182}$Os.
The concentrations to the different orbits in $^{126}$Ba and $^{124}$Xe
comes from the fact that the aligned quasiparticles
in the vacuum configurations are protons in the former
and neutrons in the latter, and then clearly explains
why the $M1$ transition is an order of magnitude stronger
in $^{126}$Ba than in $^{124}$Xe
(note also that the amount of concentration in the former is
a little bit stronger than in the latter).  In fact the value
$B(M1:I_{s\gamma}\to (I-1)_s)=0.28$ $\mu_{\rm N}^2 = 0.15$ W.u.
calculated at $\hbar\omega_\rot = 0.4$ MeV (c.f. Fig.13)
explains the observed branching ratio very well.
We think, therefore, these $^{126}$Ba data are the first evidence of the
rotationally induced "collective" $M1$ transition
between the $\gamma$- and yrast bands in the sense that it has
much larger $M1$ transitions than $E2$ transitions
though it is an even-even nucleus.

  This $M1$ matrix element on the other hand gives a possible explanation
why the linking transition between the $s\gamma$ and $g\gamma$ bands
in the $\alpha=1$ sector was not observed.
The calculated reduced transition rates give
$T(M1)=0.23\times10^{13}s^{-1}$ and $T(E2)=0.015\times10^{13}s^{-1}$
for $15_{s\gamma}\to 14_s$ ($E_\gamma=0.781$ MeV)
while $T(E2) < 0.005\times10^{13}s^{-1}$\footnote{*)}{
  This upper limit is estimated assuming a pure rotational $E2$ transitions
  as if no band crossing would exist.}
for $15_{s\gamma}\to 13_{g\gamma}$ ($E_\gamma=0.354$ MeV).
The latter should be multiplied by a hindrance factor due to the band
crossing in reality.
Consequently the intensity of the $s\gamma$ band flows
to the $s$-band rather than the $g\gamma$ band.
The two mechanisms contribute to this results: the small transition
energy in $I_{s\gamma} \to (I-2)_{g\gamma}$ and the large $M1$ transitions
in $I_{s\gamma} \to (I-1)_s$.
In the case of $^{182}$Os with negative $\gamma$ deformation, however,
this mechanism does not seem to apply because
$E_\gamma(9_{s\gamma}\to 8_s)$ is also small and the $M1$ transition
are hindered.

\vskip 3 mm
\leftline{ \it \S5-5 Effects on properties of odd-$A$ nuclei }
\vskip 2 mm

  As has been discussed in the previous sections, the character
of the lowest lying collective excitations in the $\alpha=1$ sector
is $\gamma$-vibration like below the first band crossing while it sometimes
acquires appreciable $K=1$ quadrupole strength and therefore becomes wobbling
like above the band crossing.
Rotational bands based on high-$j$ one-quasiparticle configurations
in odd-$A$ nuclei decouple into a pair of $\Delta I=2$ sequences
labeled by the different signature quantum number.
The vibrational excitation mode under consideration in even-even nuclei
affects signature-dependent properties of adjacent odd-$A$ nuclei
through the particle-collective coupling\findref\MSM.
We briefly summarize here the basic consequences of
the dynamic coupling effects.

   $B(E2:I\to I-1)$ connecting these two sequences has been known
to show signature dependence, i.e., zigzag behavior as a function of spin,
which has been observed experimentally
in some one-quasiparticle bands\findrefs\mreff\mrefg\endrefs
before the band crossing.  The theoretical studies has revealed
that the signature dependence is brought about
not only by the static\findref\mrefh but also
by the dynamic\findrefs\mrefi\mrefj\mrefk\endrefs triaxial deformations.
These two effects usually contribute in opposite sign, and, for example,
the effect of the dynamic one is partly canceled by
that of the static one in $^{157}$Ho\findref\mreff,
while the latter is dominant in $^{161}$Dy\findref\mrefg.
This dynamic effect is nothing but the manifestation of
the $K=2$ collectivity of the vibrational excitation considered
in this paper and plays an important role
to understand the electromagnetic properties of odd-$A$ nuclei.

   Above the band crossing, i.e. in three-quasiparticle bands
(e.g. $(\pi h_{11/2})^1(\nu i_{13/2})^2$) in odd-$A$ nuclei,
the signature splitting of the excitation energy between the two sequences
often becomes smaller than before the crossing and even becomes inverted
in some cases; the so-called signature-inversion.
This phenomenon has been explained in a similar way to
that for odd-odd nuclei\findrefs\mrefl\mrefm\mrefn\mrefo\mrefp\endrefs
as a result of positive $\gamma$ deformation.
But this mechanism is not appropriate to explain the inversion in
the three-quasiparticle bands which has been supposed to have
negative-$\gamma$ deformation, for example,
in the cases of $^{165, 167}$Lu\findref\mrefq.
One of the present authors showed that the phase rule, eq.(5.1), results in
stronger particle-collective coupling in the unfavoured ($I=j+odd$) sector
than in the favored ($I=j+even$) sector in negative-$\gamma$ nuclei
and consequently causes signature inversion\findref\mrefr.
In this model the effect of the static triaxiality
is usually stronger and therefore the explanation
in Refs.\findrefs\mrefl\mrefm\mrefn\mrefo\mrefp\endrefs
consistently survives for positive-$\gamma$ nuclei.
This example clearly shows, again, the importance of the collective
vibrational modes in the high-spin spectroscopy.

\vfill
\eject

\vskip 5 mm
\leftline{ \bf \S6. Concluding Remarks }
%\leftline{ \bf \S6. Summary }
%\vskip 2 mm

  We have studied the nuclear wobbling motion from a microscopic viewpoint
with paying special attention to the electromagnetic transition properties,
especially $E2$ and $M1$, and to their implications in relation to the
macroscopic Bohr-Mottelson model.
Our basic stand point is the microscopic RPA theory suitable
for high-spin states, which has been proposed by
Marshalek\findrefs\Mara\Marb\endrefs
and developed by the present authors\findrefs\SMa\MSM\endrefs.
In order to check the reliability of the microscopic framework,
the rotational perturbation to the $E2$ transitions
between the $\gamma$-vibrational and the yrast bands
in the low-spin region has been examined.
The Bohr-Mottelson's generalized intensity relation
are naturally comes out with the intrinsic transition moments
which can now be calculated microscopically.
The results of the calculations
show both the absolute magnitude and the relative phase of these moments
are well reproduced for some typical rare earth nuclei.

  The strong rotational perturbation at high-spin states
causes the structural change of the vacuum configuration, i.e.
the quasiparticle alignments or the band-crossings.
Accordingly the vibrational excitation modes are naturally
expected to change their characters.  In order to understand
the character change, the interpretations of the observable
quantities are necessary.  For this purpose the microscopic
wobbling model\findref\Mara based on the RPA theory
has been reinvestigated in the light of the macroscopic
wobbling model\findref\BMa.
It has been shown that the expressions
not only of excitation energy\findref\Mara but also of
the electromagnetic transition rates, especially the $\Delta I = \pm 1$
$E2$ transitions, can be cast into the form similar to
those given in the macroscopic rotor model.
Moreover, besides the collectivity of the RPA solutions,
a criterion to interpret the wobbling motion has been clarified:
The relative sign of the dynamic quadrupole asymmetries
around the axes perpendicular to the main rotation axis,
i.e. the transition amplitudes ${\cal Q}_y(n)$ and ${\cal Q}_z(n)$,
eqs.(4.1) and (4.3), has to be the same as that of the static asymmetry
around each axis, $\alpha_y$ and $\alpha_z$, eq.(3.8).
Just as in the rotor model, the $E2$ transitions depend also
on the three moments of inertia which determine the excitation
energy of the wobbling band at the same time.
Therefore the wobbling motion reflects
the three dimensional nature of the nuclear collective rotations.

  The wobbling motion is an unique rotational motion in the sense
that the rotation axis deviates from the inertia axis of the nuclear body.
Recently, similar rotation scheme, called "tilted axis cranking"\findref\Fraa,
is proposed and has been applied to the study of the rotational bands
with no-signature splitting.
Although both rotations are of the three dimensional nature,
they are conceptually different: the wobbling motion
is not the stationary motion and the angular momentum
and the angular frequency vectors draw different trajectories
in the body-fixed frame, while the tiled cranking is
the stationary rotation and then the angular momentum and
angular frequency vectors are parallel each other with fixed
direction with respect to the body-fixed frame.
In fact, these two vector have a definite relationship
in the wobbling motion, which is nothing but determined
by the three moments of inertia
(generally the moment of inertia tensor\findref\Godm).
Moreover, when quantized spectra are considered,
the tilted axis cranking gives a description for
an isolated band just like the usual cranking does,
but the wobbling motion as a whole corresponds to a multiple band structure,
although we concentrated mainly on the first excited band in this paper.

  Some examples of the realistic calculations are also presented
in this paper.
The sign relation mentioned above leads characteristic zigzag behaviours
in the $E2$ and $M1$ transition probabilities
as functions of the rotational frequency, depending on equilibrium
values of $\gamma$ deformation in the $(\epsilon_2,\gamma)$-plane.
This is quite analogous to the case of the $B(M1)$ between the signature
partner bands in odd-$A$ nuclei.
An important merit of the microscopic RPA formalism is that the three
moments of inertia can be easily calculated, which is highly nontrivial.
Their behaviour in three sectors of the $(\epsilon_2,\gamma)$-plane
has been calculated in a rather simplified manner
with neglecting selfconsistency of the mean-field potential.
The results clearly show that their $\gamma$ dependence
is neither irrotational nor rigid-body like.

  As a candidate of the nuclear wobbling motion which might be
identified experimentally, we have also investigated
the high-spin continuations of the so-called $\gamma$-vibrational
bands with odd-spin in $^{182}$Os, $^{124}$Xe, and $^{126}$Ba.
The selfconsistent triaxiality deformation is somewhat different
in each nucleus, and therefore the characteristic features of the
collective vibrational modes are shown also to be different.
Properties of the second-lowest RPA eigenmode, which is in some cases
also rather collective but does not satisfy the criterion
of the wobbling motion, are also discussed.

  Throughout this paper the $M1$ transition between the wobbling and yrast
bands has also been kept in scope.
It should, however, be noticed that the expression of the $B(M1)$
in the rotor model are not justified from the microscopic viewpoint.
This is because the individual property of quasiparticle orbits are essential
to understand the $M1$ transitions, which is not taken into account
in the macroscopic model, in contrast to the case of
the $E2$ transitions, where the geometrical shape of the nucleus as a whole
is mainly responsible.
Although the $M1$ transitions from the quadrupole vibrational band
are generally expected to be small at low-spin, it has been shown
that this is not always the case at higher-spin region,
where the rotation-aligned quasiparticles present in the vacuum.
Note that the $M1$ transition is only possible if there exist
the rotationally induced $K$ mixing in the RPA phonon amplitudes.
As a conspicuous example the high-spin continuation of the odd-spin
$\gamma$-vibrational band in $^{126}$Ba is predicted to have
much stronger $M1$ transition probability than the $\Delta I = \pm 1$ $E2$,
which is consistent with the observed branching ratio.
This strong $M1$ transition, on the other hand, gives a possible reason why
the $\Delta I=-2$ transition that links above and below the band crossing
in the odd-spin $\gamma$ band has not been observed.

  It is worthwhile to stress the importance of
the collective vibrational motions of either the wobbling like
or of the $\gamma$-vibrational character for the understanding
the electromagnetic property of the odd-$A$ nuclei.
The characteristic property of the vibrational modes reflect strongly
on the transition rates through the particle-vibration coupling effects,
and therefore the study of the odd-$A$ nuclei gives a valuable
testing ground to clarify the character change of the vibrational mode
predicted in the present work.  We hope that new generation
of the large array of the crystal ball will provide us more detailed
information of the electromagnetic transitions, which is necessary
to confirm the predictions and to identify the nuclear wobbling motion
if they exist.

\vskip 4 mm
\leftline{ \bf Acknowledgements }
\vskip 2 mm

Discussions with R.~Wyss are greatly acknowledged.
Early stage of this work has been done as a part of the research project on
{\it Nonlinear Dynamics of Nuclear Collective Motions}, which were
organized at the Yukawa Institute, Kyoto, in 1988-1992.
We had stimulating discussions with all the members.
This work is financially supported in part by the Grant-in-Aid for
Scientific Research from Ministry of Education, Science and Culture
(No. 06234208).

\vfill
\eject

%----------- Tables ------------------------------
%
% Table 1
%-----------------------------------------------------------------------
%\baselineskip 12 truept
%\parindent=0 truept
\vskip 5 true mm
Table 1.

{\leftskip 1 truecm \rightskip 1 truecm
  The first order coefficient of the rotational frequency dependence
calculated by the RPA theory, $a_\gamma$ in eq.(2.5) and
the extracted intensity relation parameters, $\cQ_t$ and $q$ in
eq.(2.9b).  Here ($g$) and ($\gamma$) attached to the
theoretical $q$ values mean that they are obtained by eq.(2.9b)
using experimental moment of inertia
of the ground band ($\cJ \equiv \cJ_g$)
and of the $\gamma$ band ($\cJ \equiv \cJ_\gamma$).
\par }

$$
\vbox{
\offinterlineskip
\halign{\strut
 \quad \hfil # \hfil & \quad
       \hfil # \hfil &
       \hfil # \hfil &
       \hfil # \hfil &
       \hfil # \hfil &
       \hfil # \hfil &
       \hfil # \hfil  \cr
\noalign{\hrule}
  Nucl.     &  $a_\gamma/\hbar$ [MeV$^{-1}$]   &
         $\cQ_t^{(th)}$ [eb] &  $\cQ_t^{(exp)}$ [eb] &
         $q^{(th)}(g)$ & $q^{(th)}(\gamma)$ & $q^{(exp)}$ \cr
\noalign{\hrule}
 $^{162}$Dy & 2.0  & 0.37 & 0.35$^{1)}$ & 0.027 & 0.024 &   $-$        \cr
 $^{164}$Dy & 1.9  & 0.38 & 0.34$^{2)}$ & 0.023 & 0.020 & 0.021$^{2)}$ \cr
 $^{164}$Er & 2.0  & 0.41 & 0.38$^{3)}$ & 0.031 & 0.027 &   $-$        \cr
 $^{166}$Er & 1.8  & 0.41 & 0.42$^{4)}$ & 0.024 & 0.021 & 0.022$^{4)}$ \cr
 $^{168}$Er & 1.3  & 0.42 & 0.37$^{5)}$ & 0.017 & 0.015 & 0.018$^{5)}$ \cr
\noalign{\hrule}
}
}
$$

{\leftskip 1 truecm \rightskip 1 truecm
\item{1)}
 Extracted from the $B(E2:0_g^+ \to 2_\gamma^+)$\findref\HelDy.

\item{2)}
 Extracted from the $\chi^2$ fitting of the matrix elements
    ${\cal M}(E2;\Delta I = 0, -2)_{\gamma \to g}$
        ($I_\gamma=2,4,...,10$) obtained
  by the Coulomb excitation experiment of Ref.\findref\KusaDy.
 There is an appreciable scattering of data from the relation, eq.(2.9a).
 The weights inversely propotional to the square of error bars of data
 are used for the $\chi^2$ fitting.

\item{3)}
 Extracted from the $B(E2:0_g^+ \to 2_\gamma^+)$\findref\RonEr.

\item{4)}
 {}From Ref.\findref\BMb.

\item{5)}
 Extracted from the $\chi^2$ fitting of the matrix elements
    ${\cal M}(E2;\Delta I = 0, \pm 1, \pm 2)_{\gamma \to g}$
        ($I_\gamma=2,3,...,8$) obtained
   by the Coulomb excitation experiment in Ref.\findref\KotEr
  just in the same way as 2).
\par }

%-----------------------------------------------------------------------
%\parindent=16 truept
%\baselineskip 15 truept
%

\vfill
\eject

%----------- Figure Captions ----------------------
\vskip 1cm
\centerline{\bf Figure Captions}
\vskip 3mm
\parindent 20pt
\baselineskip 18 truept

\item{Fig.1}
  The electric $E2$ transition amplitudes, eq.(2.4),
  in $^{164}$Er microscopically
  calculated by the RPA formalism as functions of the rotational
  frequency.

\item{Fig.2}
  Schematic figure depicting the analogy of the $\Delta I = \pm 1$
  $E2$ transitions between the yrast and the wobbling bands in even-even nuclei
  with the $M1$ transitions between the signature partner bands in odd nuclei.

\item{Fig.3}
 Schematic figure depicting
 the relation between the triaxiality of the mean-field and
 the $\dI = \pm 1$ $E2$ transitions from the wobbling band to
 the vacuum band.  The $(\epsilon_2,\gamma)$ plane is classified
 into three regions, for which characteristic zigzag behaviours
 of $B(E2)_{\Delta I =\pm 1}$ are shown in the middle panels.
 The transitions with stronger $B(E2)$'s
 are marked in the spectra (right panels).

\item{Fig.4}
  The $E2$ transition amplitudes, $\cQ_y$ and $\cQ_z$ (left panel) and
  $\Delta I = \pm 1$  $B(E2)$ (right panel)
  for the lowest RPA solutions with signature $\alpha=1$
  as functions of the rotational frequency in $^{182}$Os.
  The calculational procedures are the same as those in \S2 but
  the mean-field parameters are fixed as
  $\epsilon_2=0.21$, and $\gamma=-18^\circ$, and
  $\Delta_\nu=0.74$, $\Delta_\pi=0.98$ MeV,
  which are appropriate for the s-bands in $^{182}$Os\findref\Mat.
  As a reference the Weisskopf unit of $B(E2)$
  in this case is 0.0061 e$^2$b$^2$.

\item{Fig.5}
  The effective moment of inertia as functions of the triaxial parameter
  $\gamma$ at $\hbar\omega_\rot = 0.25$ MeV.
  The results for $-120^\circ < \gamma < -60^\circ$ in $^{176}$Hf,
  for $-60^\circ < \gamma < 0^\circ$ in $^{182}$Os and
  for $0^\circ < \gamma < 60^\circ$ in $^{148}$Gd are gathered in one panel.
  The mean-field parameters except $\gamma$ are fixed as
  $\epsilon_2=0.27$ and $0.22$ for $^{176}$Hf and $^{148}$Gd, respectively,
  and $\Delta_\nu = 0.6$, $\Delta_\pi = 0.6$ MeV for both nuclei.
  The same parameters are used for $^{182}$Os as Fig.4.
  Note that $^{176}$Hf and $^{148}$Gd are known to be the nuclei with
  prolate and oblate non-collective rotations so that the selfconsistent
  $\gamma$ deformations in these nuclei are $\gamma=-120^\circ$ and $60^\circ$,
  respectively.  The diabatic configurations of the vacuum states are chosen
  to be two-quasineutron four-quasiproton states for $^{176}$Hf,
  two-quasineutron states ($s$-band) for $^{182}$Os,
  and two-quasineutron and two-quasiproton states for $^{148}$Gd.
  For $^{176}$Hf and $^{148}$Gd, these configurations correspond
  to the high-$K$ states with spin $20^+$ and $18^+$ $\hbar$, respectively,
  which are the yrast with $(\pi,\alpha)=(+,0)$ for both neutron and proton
  at their selfconsistent deformations.  Unfortunately these high-$K$
  states do not coincide with any of the observed yrast isomers.

%\FIG\mfiga
\item{Fig.6}
  Routhians for the lowest lying state in the
  $\alpha=1$ sector as functions of the rotational frequency in $^{124}$Xe.
  The calculated one is shown by the solid line and
  the experimental one, which is the routhian of the odd-spin members of
  the so-called $\gamma$ band, by the dashed line, and
  the observed yrast band by the short-dashed line.
  Experimental routhians are shown with respect to
  a reference band with $\cJ_0=15.0\hbar^2/$MeV,
  $\cJ_1=25.0\hbar^4/{\rm MeV}^3$, $e_s=3.03$ MeV and $i_s=7.14\hbar$.
  Parameters used for the calculation are $\epsilon_2=$0.19,
  $\gamma=-45^\circ$ and $\Delta _\nu = 0.9$, $\Delta _\pi = 1.1$ MeV.

%\FIG\mfigb
\item{Fig.7}
  Calculated effective moments of inertia of the lowest lying state in
  the $\alpha=1$ sector in $^{124}$Xe as
  functions of the rotational frequency.
  Parameters used are the same as in Fig.6.

\item{Fig.8}
  The $E2$ transition amplitudes, $\cQ_y$ and $\cQ_z$ (left panel) and
  $\Delta I = \pm 1$  $B(E2)$ (right panel)
  for the lowest RPA solutions with signature $\alpha=1$
  as functions of the rotational frequency in $^{124}$Xe.
  Parameters used are the same as in Fig.6.
  As a reference the Weisskopf unit of $B(E2)$
  in this case is 0.0037 e$^2$b$^2$.

\item{Fig.9}
  The $M1$ transition amplitudes, $i\mu_y$ and $\mu_z$ (left panel) and
  $B(M1)$ (right panel) for the lowest RPA solutions with signature $\alpha=1$
  as functions of the rotational frequency in $^{124}$Xe.
  Parameters used are the same as in Fig.6.
  As a reference the Weisskopf unit of $B(M1)$ is 1.79 $\mu_N^2$.
  Note that the relative phase between the $M1$ amplitudes
  and the $E2$ amplitudes (Fig.8) is meaningful
  and gives the sign of the $E2/M1$ mixing ratio.

%\FIG\mfigc,\mfigd
\item{Fig.10}
  Excitation energies (left panel) and the $E2$ transition amplitudes
  (right panel) of the lowest and second-lowest states in the
  $\alpha=1$ sector in $^{124}$Xe calculated at
  $\hbar\omega_\rot=0.3$ MeV as functions of $\gamma$.
  Parameters used for the calculation except $\gamma$ are the same as Fig.6.

%\FIG\mfigf
\item{Fig.11}
  The same as Fig.6 but for $^{126}$Ba.
  Parameters used for the calculation are $\epsilon_2=$0.24, $\gamma=$0,
  and selfconsistently calculated pairing gaps are
  $1.16 \geq \Delta_\nu \geq 1.03$ MeV and
  $0.76 \geq \Delta_\pi \geq 0.71$ MeV for
  $0.4 \leq \hbar\omega_\rot \leq 0.5$ MeV.
  A reference band with $\cJ_0=14.0 \hbar^2/$MeV,
  $\cJ_1=31.0 \hbar^4/{\rm MeV}^3$, $e_s$ = 3.00 MeV and $i_s=7.64\hbar$
  is used here.
  Note that the $\gamma$-ray which links above and below the band
  crossing has not been observed (see the text).

\item{Fig.12}
  The same as Fig.8 but for $^{126}$Ba.
  Parameters used are the same as in Fig.11.
  As a reference the Weisskopf unit of $B(E2)$
  in this case is 0.0038 e$^2$b$^2$.

\item{Fig.13}
  The same as Fig.9 but for $^{126}$Ba.
  Parameters used are the same as in Fig.11.

\vfill
\eject

%----------- references  ----------------------

\outrefs

\end